\documentclass[12pt,subeqn,a4paper]{article}
\usepackage{amsmath}
\usepackage{amssymb}
\usepackage{mathtools}
\usepackage{graphicx}
\usepackage{geometry}
\usepackage{caption}
\usepackage{float}
\usepackage{subcaption}
\usepackage{hyperref}
\usepackage{tocloft}
\usepackage{verbatim}
\usepackage{comment}
\usepackage{titlesec}
\usepackage{slashed}
\usepackage{bm}
\hypersetup{colorlinks=true,linkcolor=blue}
\newcommand{\Z}{\mathbf{Z}}
\newcommand{\R}{\mathbf{R}} 
\newcommand{\C}{\mathbf{C}} 

\newcommand{\ch}{\text{ch}} 

\newcommand{\Tr}{\text{Tr}}

\newcommand{\Sym}{\text{Sym}}
\newcommand{\Hom}{\text{\Hom}}

\newcommand{\diag}{\text{diag}}

\DeclareMathSymbol{:}{\mathord}{operators}{"3A}

\DeclarePairedDelimiter\bra{\langle}{\rvert}
\DeclarePairedDelimiter\ket{\lvert}{\rangle}
\DeclarePairedDelimiterX\braket[2]{\langle}{\rangle}{#1
\delimsize\vert #2}
\DeclarePairedDelimiterX\ketbra[2]{\delimsize\vert}{\delimsize\vert}{#1
\rangle\langle
 #2}
 
\begin{document}

\begin{titlepage}
\begin{flushright}
{\bf November 2023} \\ 
\end{flushright}
\begin{centering}
\vspace{.2in}
 {\large {\bf Superconformal Quantum Mechanics and Growth of Sheaf Cohomology}}

\vspace{.3in}

Nick Dorey, Boan Zhao\\
\vspace{.1 in}
DAMTP, Centre for Mathematical Sciences \\ 
University of Cambridge, Wilberforce Road \\ 
Cambridge CB3 0WA, UK \\
{\tt N.Dorey@damtp.cam.ac.uk, bz258@cam.ac.uk} \\
\vspace{.2in}
%
%
\vspace{.4in}
{\bf Abstract} \\

We give a geometric interpretation for superconformal quantum mechanics defined on a hyper-K\"{a}hler cone which has an equivariant symplectic resolution. BPS states are identified with certain twisted Dolbeault cohomology classes on the resolved space and their index degeneracies can also be related to the Euler characteristic computed in equivariant sheaf cohomology. In the special case of the Hilbert scheme of $K$ points on $\mathbb{C}^{2}$, we obtain a rigorous estimate for the exponential growth of the index degeneracies of BPS states as $K\rightarrow \infty$. This growth serves as a toy model for our recently proposed duality between a seven dimensional black hole and superconformal quantum mechanics. 
\end{centering}


\end{titlepage}

\setcounter{tocdepth}{2}


\vspace{2em}

\section*{Introduction}

Supersymmetric quantum mechanics on a compact Riemannian manifold $\mathcal{X}$ has a beautiful geometric interpretation \cite{Witten} in which supersymmetric ground states correspond to de Rham cohomology classes and the Witten index coincides with the topological Euler character $\chi(\mathcal{X})$. In this paper we will present a comparable geometric interpretation for the index of {\em superconformal} quantum mechanics (SCQM) \cite{Eliezer} (See also \cite{Ivanov} and references therein). 
If $M$ is a hyper-K\"{a}hler cone, supersymmetric quantum mechanics on $M$ admits an $\mathfrak{osp}(4^{*}|4)$ superconformal algebra \cite{Singleton1}. Assuming a spectrum consisting of unitary irreps of this algebra, one can define a superconformal index which counts (semi-)short multiplets \cite{SingletonDorey}. These special irreps contain BPS states which saturate a Bogomol'nyi bound. However, this definition is somewhat formal due to the singularity at the tip of the cone. To make progress it is necessary to regulate the quantum mechanics by resolving this singularity. For this reason we focus on hyper-K\"{a}hler cones $M$ which have an {\em equivariant symplectic resolution}\footnote{For a review of the geometry of these spaces see section 2 of \cite{braden2012}.} $\tilde{M}\to M$. A large class of such spaces is provided by Nakajima quiver varieties. These spaces can be constructed as hyper-K\"{a}hler quotients and can also be realised as complex algebraic varieties. The resolution is typically characterised by one or more real parameters $\xi_{\mathbb{R}}$ with superconformal invariance recovered in the limit $\xi_{\mathbb{R}}\rightarrow 0$. .
\paragraph{}
In the first part of this paper, we will give an explicit definition for the superconformal index. In particular, we study a certain supersymmetric quantum mechanics denoted $\widetilde{SQM}$, defined on the resolved space $\tilde{M}$ which reduces to SCQM on $M$ as the resolution parameters are take to zero. We define a Witten index for $\widetilde{SQM}$ which counts      
BPS states on $\tilde{M}$ and compute it using standard localisation methods. The index of $\widetilde{SQM}$ turns out to be independent of the resolution parameters and provides a sensible definition of the superconformal index. Consistency of this definition with  the $\mathfrak{osp}(4^{*}|4)$ symmetry of the conformal point is not immediate, but we will prove two necessary conditions for this below. The index on $\tilde{M}$ counts BPS states annihilated by $Q,Q^\dagger$ where $Q$ is the Dolbeault operator twisted by the moment map of a certain holomorphic Killing vector on $\tilde{M}$. Direct computation of the index using localisation shows that it equals the equivariant $\chi_y$ genus of $\tilde{M}$ as proposed in \cite{DoreyBarnsGraham}. For projective manifolds, the BPS states defined by $Q$ naturally biject with harmonics forms of degree $p,q$ which also naturally biject with $H^q(\Omega^p)$ taken in the Zariski topology. Hence the agreement between the corresponding indices in that case can be explained by this bijection. Although we do not know of a proof, it seems possible that such a bijection exists for the class of non-compact manifolds considered in this paper.
\paragraph{} 
In the special case where $M=\mathcal{M}_{K,N}$, the moduli space of $SU(N)$ Yang-Mills $K$-instantons with a partial Uhlenbeck compactification, these BPS states can be provisionally identified with the microstates of a supersymmetric AdS pp-wave black hole constructed in \cite{NickRishiAndy}. For $N>>1$, the asymptotic growth of the index degeneracy with $K$ and other charges reproduces the Bekenstein-Hawking entropy of the black hole. In the second part of the paper we estimate this growth rigorously for the special case $N=1$ where the resolved space $\tilde{M}$ coincides with the Hilbert scheme of $K$ points on $\mathbb{C}^{2}$. Although not directly related to gravity, this simple model exhibits an exponential growth in the index degeneracy of states which is quite similar to the case of general $N$. It therefore provides a useful toy model in which we can test the methods used in  \cite{NickRishiAndy}. {Moreover, the technique used in the proof works for a large class of examples corresponding to plethystic exponentials of rational functions. The authors hope it will lead to more rigorous results in the theory of asymptotics. Previous works related to the asymptotic of the index includes  \cite{LeeNahmgoong}\cite{kim2017asymptoticM5brane}}. In the remainder of this section we will briefly state the main results of the paper with details relegated to the subsequent sections.                
\paragraph{}
Any hyper-K\"{a}hler cone $M$ is equipped with a triholomorphic homothetic vector $D$ whose Lie derivative serves as the dilation operator in superconformal quantum mechanics (SCQM) on $M$. We choose a preferred complex structure $I$ and the vector field $V = ID$ is a holomorphic killing vector field under which the holomorphic symplectic form coming from the hyperkahler structure has weight two. As we review below, the statement about $V$ is also true for the equivariant symplectic resolution $\tilde{M}$. These conditions allow us to write down a Lagrangian \cite{AlvarezGaumePotential} for particle motion on $\tilde{M}$ with $\mathcal{N}=(2,2)$ supersymmetry with in which the norm of $V$ (which equals norm of $D$) appears as a potential\footnote{Here $\phi$ and $\psi$ denote bosonic and fermionic degrees of freedom on $\tilde{M}$. 
See section 2 for notation and further details.}:    
\begin{equation}\label{intro:lagrangian_with_potential}
    L(\phi, \psi, \bar{\psi}) =\frac{1}{2}(\dot{\phi},\dot{\phi}) + (\bar{\psi},\dot{\psi}) -\frac{1}{4} R(\psi,\psi,\bar{\psi},\bar{\psi}) - \frac{1}{2}(D, D) - i \nabla_iD_j \psi^i\bar{\psi}^j
\end{equation}
In the general setting described above, the potential term in the Lagrangian grows asymptotically in all directions. Thus, although $\tilde{M}$ is non-compact, the resulting supersymmetric quantum mechanics, denoted $\widetilde{SQM}$, has a discrete energy spectrum bounded below by zero.
\paragraph{}
States in $\widetilde{SQM}$ are represented by differential forms on $\tilde{M}$. They carry a conserved charge $Q_{V}$ corresponding to the eigenvalue of $-i\mathcal{L}_{V}$ acting on forms. All states obey a Bogomol'nyi bound $E\geq |Q_{V}|$ where $E$ is the eigenvalue of the Hamiltonian. The so-called BPS states saturating this bound are annihilated by $Q,Q^\dagger$ where    
\begin{equation*}
Q = \bar{\partial} + \bar{\partial} C\wedge
\end{equation*}
is a twisted Dolbeault operator and $C$ denotes the moment map of $V$, defined up to a constant. Assuming that the simuiltaneous eigenspaces of $\{Q,Q^\dagger\}$ and $Q_V$ in the full Hilbert space are all finite dimensional, one can show using a hodge-type decomposition that the BPS states biject with Q-cohomology classes, provided we restrict to states with definite quantum number under $Q_V$. Note that that the $\mathfrak{osp}(4^{*}|4)$ superconformal invariance of SCQM is broken to a conventional $\mathcal{N}=(2,2)$ supersymmetry algebra in $\widetilde{SQM}$. The Bogomol'nyi bound discussed here is the one corresponding to the latter algebra. Nevertheless both the bound and the corresponding BPS states reduce to those of the full superconformal algebra as the resolution parameters vanish.  
\paragraph{}
As usual we can define an index which only receives contributions from BPS states, 
\begin{equation}\label{intro:index_y}
    Z[t,y,z_i] = \Tr((-1)^F\exp(-\mu\{Q, Q^\dagger\})t^{L_t}y^{L_y}z_i^{L_i}), \mu>0
\end{equation}
Here $z_i$ are the fugacities corresponding to the additional holomorphic killing vector fields $Z_i$ which generate $U(1)$ isometries on $\tilde{M}$. The product over $i$ in the trace is implicit.
The conserved charges are defined as follows:
\begin{equation*}
    F = p + q - \frac{\dim_\C \tilde{M}}{2}, L_t = -i\mathcal{L}_{V} - (p - \frac{\dim_\C \tilde{M}}{2}), L_y = p - \frac{\dim_\C\tilde{M}}{2}, L_i = -i\mathcal{L}_{Z_i}
\end{equation*}
Here $p$ and $q$ respectively denote the holomorphic and anti-holomorphic degrees of differential forms. 
\paragraph{}
Using standard techniques of supersymmetric quantum mechanics we can localise the index to the fixed points of a generic group action on $\tilde{M}$. The resulting formula,  
\begin{equation}\label{intro:index_abstract}
Z[t,y,z_i]  = \sum_P (-\hat{y})^{-\dim_\C \tilde{M}/2}\prod_i \frac{1 - \hat{y}\chi_i^{-1}(P)}{1 - \chi_i^{-1}(P)}
\end{equation}
with $\hat{y}:=y/t$,  expresses the index as a sum over fixed points. The contribution of each fixed point $P$ is given in terms of the one dimensional characters $\chi_i(P),i=1,...,\dim_\C\tilde{M}$ ({depending} on $t$ and the $z_{i}$) of the group action on the tangent space to the manifold at $P$. As we discuss further below, this formula for the index agrees with equivariant $\chi_y$ genus of $\tilde{M}$. 
\paragraph{}
By construction, the index $Z[t,y,..]$ provides a regularised version of the superconformal index. However, such an interpretation requires that the index takes the form of a sum over the contributions of {irreducible} representations of $\mathfrak{osp}(4^*|4)$ superconformal {algebra \cite{SingletonDorey}}.
This in turn demands several non-trivial properties of of the index which are not immediately obvious from the definition or the localisation formula. In particular the index must be invariant under $y\rightarrow 1/y$ and must have a finite limit as $t\rightarrow 0$ and a Taylor expansion in nonnegative powers of $t$. In Section 2, we prove these properties directly from the geometric definition of the index.
\paragraph{}
In the remaining sections of the paper we analyse the special case where $\tilde{M}$ is the Hilbert scheme of $K$ points on $\mathbb{C}^{2}$ which coincides with the moduli space of $K$ non-commutative abelian instantons $\mathcal{M}_{K,1}$. This space has an additional $U(1)$ holomorphic isometry with charge $L_{x}$ which can be used to grade the index. Our analysis in the general case guarantees the index has a Taylor expansion of the form,   
\begin{equation*}
    Z_K(t,x,y) = \sum_{L_t\geq 0,  L_y, L_x}\mathcal{C}(L_t,L_y,L_x,K)t^{L_t}y^{L_y}x^{L_x}
\end{equation*}
and that the coefficents $\mathcal{C}\in\mathbb{Z}$ appearing in this expansion correspond to an a certain alternating sum of BPS states. It also corresponds to a certain alternating sum of sheaf cohomology classes with definite grade:
\begin{equation*}
    \mathcal{C}(L_t,L_y, L_x,K) = \sum_{q\geq 0}(-1)^q d(q,L_t + L_y,L_x,K)
\end{equation*}
where $d(q,L_t+L_y,L_x,K)$ denotes the dimension of the $t^{L_t + L_y}x^{L_x}$ weight space of $H^q(\Omega^p, \tilde{M}=\mathcal{M}_{K,1}),p = L_y + \dim_\C\tilde{M}/2$. Later in the paper $Z_K$ and coefficient $\mathcal{C}$ differ from these definition by an overall sign. Since we are only interested in the growth of $|\mathcal{C}|$ the overall sign is not a problem.

\paragraph{}
In recent work \cite{NickRishiAndy} we studied the growth of the index coefficients $\mathcal{C}$ for $X=\mathcal{M}_{K,N}$ using electromagnetic duality and other physics input. Our main result was 
an asymptotic formula for the exponential growth of these coefficents. In this paper we give a rigorous derivation of this asymptotic in the $N=1$ case where the relevant moduli space is the Hilbert scheme described above. In particular, we prove that when $L/K$ is fixed (and $\geq 4$),
\begin{equation}
C(L,0,0,K) = \exp(2\pi 24^{-1/4}\sqrt{L}K^{1/4}+...)(\cos(2\pi 24^{-1/4}\sqrt{L}K^{1/4}+...) + o(1)), K\to \infty
\end{equation}
where $...$ denote subleading terms\footnote{We will not compute the exact coefficients in the subleading corrections.} which are real polynomials in $K, \log(K)$ when $K/L$ is fixed. 
In this paper $A = \mathcal{O}(B)$ means there is a universal constant $C>0$ so that $|A|\leq C|B|$. $\mathcal{O}_{K/L}(B)$ means there is a constant $C$ which depends only on $K/L$ so that $|A|\leq C(K/L)|B|$.

The proof relies on an a exact resummation of the generating function, 
\begin{equation}
  \sum_{K=0}^{\infty}\, Z_{K}[t,y,x] \, q_{\tau}^K = \exp\left(\sum_{n\geq 1}\frac{1}{n}\frac{q_\tau^n}{1-q_\tau^n}\frac{\hat{y}^{-n}(1 - \hat{y}^nq_1^n)(1 - \hat{y}^nq_2^n)}{(1-q_1^n)(1 - q_2^n)}\right)
\end{equation}
which has been rigorously proven \cite{Rains}. We then extract the coefficient $\mathcal{C}$ using an contour integral, and applies the Hardy-Ramanujan saddle point\cite{HardyRamanujan} method to compute the asymptotic. We also performed numerical check of the asymptotic formula up to $L\approx K \approx 60$ and found good agreement.

\section*{Superconformal quantum mechanics on hyperkahler cones}
In this section we review superconformal quantum mechanics on general hyperkahler cones and explain the superconformal index and its computation.

A hyperkahler cone is a connected hyperkahler manifold (away from the singularities) $(M,g,I,J,K)$ with an additional triholomoprhic homothety $D$ (a vector field on $M$) such that
\begin{equation*}
    \mathcal{L}_D I = \mathcal{L}_DJ = \mathcal{L}_D K = 0, \mathcal{L}_D g = 2g
\end{equation*}
where $I,J,K$ are the three complex structures and $g$ is the hyperkahler metric \cite{DoreyBarnsGraham}. $\mathcal{L}$ denotes lie derivative. We further assume that $ID, JD, KD$ generate $U(1)$ isometries in the sense that the flow along any of these three vector fields by $2\pi$ is the identity. We also assume the existence of a function $C: M\to \R$ so that $D^2 = 2C, D^i  = \nabla^iC$. In the simplest example of flat $\R^{4N}$ with the standard Euclidean metric, the vector field $D = x^i\partial_{x^i}$ and $C = \frac{1}{2}\sum_i |x^i|^2$ where $x^i$ are flat coordinates of $\R^{4N}$. It can be shown that a supersymmetric particle on a hyperkahler cone with Lagrangian
\begin{equation*}
    L(\phi, \psi,\bar{\psi}) = \frac{1}{2}(\dot{\phi},\dot{\phi}) + (\bar{\psi}, \dot{\psi}) - \frac{1}{4}R(\psi,\psi,\bar{\psi},\bar{\psi})
\end{equation*}
with SUSY transformation law
\begin{equation*}
    \delta \phi = \epsilon \psi + \bar{\epsilon} \bar{\psi}, \delta \psi = -\bar{\epsilon}\dot{\phi}, \delta\bar{\psi} = -\epsilon \dot{\phi}
\end{equation*}
has $\mathfrak{osp}(4^*|4)$ symmetry \cite{SingletonThesis}, where $*$ denotes a real form of $\mathfrak{osp}(4|4)$. The bosonic part of the real form is $so(2,1)\oplus su(2)\oplus so(5)$. $\phi$ is a map $\R \to M$ and $\psi$ is an odd section of the complexified pullback tangent bundle via $\phi$. $(\hspace{0.2cm},\hspace{0.2cm})$ is the Riemannian metric $g$ and dot means time derivative. The Hilbert space of the quantum mechanics consists of
complex differential forms on $M$ which decay at infinity.

Our goal is to compute a Witten type index of the form $\Tr((-1)^F\exp(-\mu \{Q,Q^\dagger\})...)$ using localisation, where $F$ is a Fermion number and $Q$ is a supercharge, both of which will be defined below. The $...$ denotes the insertion of conserved charges associated with an isometric torus action of $M$. Most hyperkahler cones are singular. As a result, localisation computation is difficult as the fixed locus of the torus action is usually singular. As a result, we consider a regularized version of the quantum mechanics {$\widetilde{SQM}$} which lives on a resolved space $\tilde{M}\to M$ where $\tilde{M}$ is smooth. We regard $M$ as a Kahler manifold by picking the complex structure $I$. The manifold $\tilde{M}$ is also Kahler and the Kahler metric on $\tilde{M}$ should asymptotically agree with the Kahler metric on $M$ near infinity. We also assume the existence of a vector field $D$ on $\tilde{M}$ so that $ID$ generates a U(1) Kahler isometry of $\tilde{M}$ and the resolution $\tilde{M}\to M$ is $U(1)$ equivariant. $D$ is usually not a homothety on $\tilde{M}$ and quantum mechanics on $\tilde{M}$ is usually not conformal. We define the function $C$ as the moment map associated with $ID$ on $\tilde{M}$: $dC(\cdot) = -\omega(ID, \cdot)$ where $\omega$ is the Kahler form of $\tilde{M}$. Equivalently, $\nabla^i C = D^i$. Hence $C$ is defined up to an overall constant. Finally, we also assume the existence of a hyperkahler structure\footnote{the map $\tilde{M}\to M$ does not need to preserve the full hyperkahler structure} on $\tilde{M}$ extending the Kahler structure. The hyperkahler structure naturally defines a holomorphic symplectic form $\omega_\C$ on $\tilde{M}$ which we assume to have weight two under $D$: $\mathcal{L}_D\omega_\C = 2\omega_{{\C}}$. {Such a resolution $\tilde{M}\to M$ is usually called an equivariant symplectic resolution. A common way of constructing such a resolution is to use twisted GIT quotient. When $M$ is an ordinary GIT quotient $X//G$ (the set of Zariski closed $G$-orbit in $X$ where $X$ an affine variety and $G$ a reductive algebraic group), we can replace the quotient by a twisted GIT quotient $X//_\chi G$, where $\chi: G\to \C$ is a character. The twisted GIT quotient is often nonsingular and there is a natural projective map $X//_\chi G\to X//G$ which resolves the singularity in $M = X//G$ \cite{nakajima1999lectures}. In this work, $M$ is a hyperkahler quotient with real FI parameters $\xi_\R$ and such resolution is the same as setting $\xi_\R$ to be nonzero.}

We study supersymmetric quantum mechanics with target space $\tilde{M}$ and regard it as a regularized version of quantum mechanics on $M$. Its classical Lagrangian is the usual Riemannian supersymmetric quantum mechanics with a superpotential \cite{AlvarezGaumePotential}
\begin{equation}\label{lagrangian_with_potential}
    L(\phi, \psi, \bar{\psi}) =\frac{1}{2}(\dot{\phi},\dot{\phi}) + (\bar{\psi},\dot{\psi}) -\frac{1}{4} R(\psi,\psi,\bar{\psi},\bar{\psi}) - \frac{1}{2}(D, D) - i \nabla_iD_j \psi^i\bar{\psi}^j
\end{equation}
with SUSY transformation law
\begin{equation*}
    \delta \phi = \epsilon\psi + \bar{\epsilon}\bar{\psi}, \delta\psi = -\bar{\epsilon}(\dot{\phi} - iD), \delta\bar{\psi} = -\epsilon(\dot{\phi} + iD)
\end{equation*}
and supercharges
\begin{equation*}
    Q^{Riem} = i(\dot{\phi} - iD, \bar{\psi}), \overline{Q^{Riem}} = -i(\dot{\phi} + iD, \psi)
\end{equation*}
Under the SUSY transformation law the Lagrangian changes by a total derivative
\begin{equation*}
    \delta L = \frac{d}{dt}(\bar{\epsilon}\bar{\psi},\dot{\phi}) - i\epsilon\frac{d}{dt}(D,\psi)
\end{equation*}
The Hilbert space is the space of smooth differential forms on $\tilde{M}$ which decay at infinity.
The Riemannian supercharge and the Hamiltonian are
\begin{equation*}
    Q^{Riem} = d + dC\wedge, H = \frac{1}{2}\{Q, Q^\dagger\} = \frac{1}{2}\{d, d^\dagger\} + \frac{1}{2}|dC|^2 + \frac{1}{2}\mathcal{L}_{D} + \frac{1}{2}\mathcal{L}_D^\dagger 
\end{equation*}
Since $V:= ID$ is a holomorphic killing vector field by assumption, the Lagrangian \eqref{lagrangian_with_potential} has a $U(1)\times U(1)$ R-symmetry and hence $(2,2)$ supersymmetry. The supercharges are
\begin{equation*}
    Q = \bar{\partial} + \bar{\partial} C\wedge, \tilde{Q} = \partial + \partial C\wedge, Q + \tilde{Q} = Q^{Riem}
\end{equation*}
And the commutators are
\begin{equation*}
    \{Q, Q^\dagger\} = H + i \mathcal{L}_{ID}, \{\tilde{Q},\tilde{Q}^\dagger\} = H - i\mathcal{L}_{ID}, \{Q,\tilde{Q}\} =\{Q, \tilde{Q}^\dagger\}=0
\end{equation*}
Since $\{Q,Q^\dagger\}\geq 0, \{\tilde{Q},\tilde{Q}^\dagger\}\geq 0$ we have the BPS bound
\begin{equation*}
    H\geq |i\mathcal{L}_V|
\end{equation*}
The Hamiltonian simplifies on the unresolved space. In this case the lie derivative term $Lie_D + Lie_D^\dagger$  can be expressed using form degree
\begin{equation*}
    \mathcal{L}_D + \mathcal{L}_D^\dagger = 2(p+q) - \frac{1}{2}\dim_\C(\tilde{M})
\end{equation*}
where $p,q$ are the holomorphic and antiholomorphic degrees.
Hence we can rewrite the BPS bound as the following:
\begin{equation*}
    \frac{1}{2}\{d,d^\dagger\} + \frac{1}{2}|dC|^2 \geq |i\mathcal{L}_{ID}| - (p+q) + \frac{1}{4}\dim_\C \tilde{M}
\end{equation*}
States which are annihilated by both $Q$ and $Q^\dagger$ are known as BPS states. We are only interested in the states with definite quantum numbers under global symmetries. Since $Q = \exp(-C)\circ \bar{\partial}\circ \exp(C)$, BPS states with antiholomorphic degree $q = 0$ are of the form $\omega \exp(-C)$ where $\omega$ is an algebraic differential forms with at most polynomial growth at infinity. The exponentially decaying factor $\exp(-C)$ ensures the normalizability of these states. Rigorous results concerning the operators $Q$ can be found in \cite{Braverman}.

Our goal is to count the alternating sum of these BPS states using a superconformal index \cite{DoreyBarnsGraham}
\begin{equation}\label{index_y}
    Z[t,y,z_i] = \Tr((-1)^F\exp(-\mu\{Q, Q^\dagger\})t^{L_t}y^{L_y}z_i^{L_i}), \mu>0
\end{equation}
Here $z_i$ are the fugacities corresponding to additional holomorphic killing vector fields $Z_i$ which generate $U(1)$ isometries on $\tilde{M}$. The product over $i$ is implicit in the index.
The conserved charges are defined as follows:
\begin{equation*}
    F = p + q - \frac{\dim_\C \tilde{M}}{2}, L_t = -i\mathcal{L}_V - (p - \frac{\dim_\C \tilde{M}}{2}), L_y = p - \frac{\dim_\C\tilde{M}}{2}, L_i = -i\mathcal{L}_{Z_i}
\end{equation*}
We assume that $t, y, z_i\in U(1)$. Later we will see that the index can be resummed to a rational function and hence the domain of $t,y,z_i$ can be enlarged. The usual argument shows that the index is independent of $\mu$ and receives contribution from {BPS states only}:
\begin{equation*}
    Z[t,y,z_i] = Tr_{{\text{BPS}}}((-1)^Ft^{L_t}y^{L_y}z_i^{L_i})
\end{equation*}

Now we discuss a symmetry of this index.
$y$ is the Cartan generator of an $su(2)$ R-symmetry: the other two generators of $su(2)$ (more precisely its complexification) are $\omega_\C\wedge, (\omega_\C\wedge)^\dagger$ with commutators
\begin{eqnarray*}
    \left[p - \dim_\C\tilde{M}/2, \omega_\C \wedge\right] = 2\omega_\C \wedge\\ 
    \left[p - \dim_\C\tilde{M}/2, (\omega_\C\wedge)^\dagger\right] = -2(\omega_\C\wedge)^\dagger\\ 
    \left[\omega_\C\wedge, (\omega_\C\wedge)^\dagger\right] = 4(p - \frac{\dim_\C \tilde{M}}{2})
\end{eqnarray*}
The last identity follows from the hyperkahler structure of the manifold.
As a result we have a unitary action of the following real lie algebra
\begin{equation*}
    \underbrace{su(2)}_{\substack{\omega_\C\wedge\\p - \dim_\C \tilde{M}/2\\(\omega_\C\wedge)^\dagger}} \oplus \underbrace{u(1)}_{\mathcal{L}_{ID}} \oplus \underbrace{u(1)}_{\mathcal{L}_{Z_1}}\oplus...
\end{equation*}
We can expand the superconformal index as a product of characters of this real lie algebra. Since the $su(2)$ characters are invariant under $y\to y^{-1}$, it follows that the index is also invariant under $y\to y^{-1}$.

Now let us derive an explicit localisation formula for the index. We introduce a new variable $\hat{y} = y/t$ which makes our future computation easier. The index written using $\hat{y}$ is
\begin{equation}\label{index_y_hat}
    Z[t,\hat{y},z_i] = \Tr((-1)^q\exp(-\mu\{Q,Q^\dagger\})t^{-i\mathcal{L}_{ID}}z_i^{-i\mathcal{L}_{Z^i}}(-\hat{y})^{p - \dim_\C/2})
\end{equation}

The superconformal index can be computed using a direct path integral manipulation as in \cite{AlvarezGaumeIndex}. Here we briefly sketch the details. First, we write the Witten index (without the insertion of the conserved charges) $\Tr((-1)^F\exp(-\mu \{Q,Q^\dagger\}))$ (formally) as a path integral (up to a sign $(-1)^{\dim_\C\tilde{M}/2}$)
\begin{equation}\label{path_integral}
    \int d\phi d\psi d\bar{\psi} \exp\left(-\int_0^\mu \tilde{L}(\phi, \psi,\bar{\psi})\right), \tilde{L} = L - \nabla_i V_j \psi^i \bar{\psi}^j
\end{equation}
with periodic boundary conditions
\begin{equation*}
    \phi(\mu) = \phi(0),\psi(\mu) = \psi(0), \bar{\psi}(\mu) = \bar{\psi}(0)
\end{equation*}

The Witten index is ill-defined in our setting since there are infinitely many states annihilated by both $Q$ and $Q^\dagger$. As a result, this path integral is ill-defined. It becomes well-defined after we insert the conserved charges. This procedure is equivalent to changing the boundary conditions of fields in the path integral: when we compute $\bra{p}\exp(-\mu H) U^{-1}\ket{p}$ for $p\in \tilde{M}$, the Euclidean transition amplitude between $U^{-1}\ket{p}$ and $\bra{p}$, we need to integrate over paths which start at $U^{-1}p$ and end at $p$. Here $U:\tilde{M}\to \tilde{M}$ is a Kahler isometry which also acts on the Hilbert space as a unitary operator. In our case $U^{-1} = t^{-i\mathcal{L}_{ID}}z_i^{-i\mathcal{L}_{Z^i}}$ When we take trace over $p$ the boundary condition becomes $\phi(\mu) = U\phi(0), \psi(\mu) = U_*\psi(0)$, {where $U_*$ is the derivative of $U$}. The power of $\hat{y}$ contains the holomorphic degree only so we {rotate} the (1,0) part of $\psi$ by $\hat{y}$:
\begin{equation*}
    \psi^{0,1}(\mu) = \psi^{0,1}(0), \psi^{1,0}(\mu) = \exp(i\gamma)\psi^{1,0}(0),\psi = \psi^{1,0} + \psi^{0,1}, \hat{y} = \exp(i\gamma)
\end{equation*}
where I have decomposed $\psi$ into its $(1,0)$ and $(0,1)$ part.
In summary{,} the {correct} boundary condition for (\ref{index_y_hat}) should be
\begin{equation*}
    \phi(\mu) = U\phi(0), \psi^{0,1}(\mu) = U_*\psi^{0,1}(0), \psi^{1,0}(\mu) = U_*\exp(i\gamma)\psi^{1,0}(0)
\end{equation*}
{Now} we take $\mu \to 0$. If a {path} $\phi$ starts at $p\in \tilde{M}$ and ends at $U p \in \tilde{M}$, the time scales as $\mu$ and so {the} velocity scales as $1/\mu$ and the action scales as velocity square times time $\mu^{-2}\mu \sim \mu^{-1}$. And hence as a long as $Up \neq p$ the action would diverge and its exponential would tend to zero. The only paths whose action {do} not diverge {are} when $Up = p$ and the path is constant. For {generic} $t,z_i$ the fixed point{s} of $U$ {agree} with the fixed points of torus action generated by $V, Z^i$. We assume that there are only finitely many fixed points. The path integral localizes to constant paths at these fixed points. We assume that it localizes to $\psi = 0$ because the SUSY variation $\delta \phi = 0$ when $\psi = 0$.

It remains to compute {the quadratic fluctuations} around these saddle points. Let $P$ be a fixed point. We diagonalize the action of $U_*$ on the tangent space {of $\tilde{M}$} at $P$:
\begin{equation*}
    U_*(T_P\tilde{M}) = \diag(\exp(i\theta_1),\exp(i\theta_2),...,\exp(i\theta_{\dim_\C\tilde{M}}))
\end{equation*}
So the tangent space is a direct sum of one dimensional complex vector spaces. We will focus on the first summand and later take product over all the summands. The index is invariant under rescaling the potential $C$ by 
$\lambda C$ {whenever} $\lambda >0$ and so we can let $\lambda \to 0$ in our computation {and discard the function $C$}. We also scale the time variable $\theta = 2\pi t/\mu$ so that $\theta\in [0,2\pi]$.

The quadratic fluctuation{s} consist
of {the} bosonic determinant $\det(-d^2/d\theta^2)$ and two fermionic determinants $\det(d/d\theta)_{0,1}$ and $\det(d/d\theta)_{1,0}$ coming from the quadratic fluctuation{s} of $\psi^{1,0}$ and $\psi^{0,1}$ respectively. The final result is (up to a constant which depends only on $\mu$):
\begin{equation}\label{quadratic_fluctuation}
   \det(d/d\theta)_{0,1}\det(d/d\theta)_{1,0} (\det(-d^2/d\theta^2))^{-1/2}
\end{equation}
To compute the bosonic determinants we use the following real eigenbasis for $-d^2/d\theta^2$: ($\phi(\theta) = P + \delta \phi(\theta$))
\begin{align*}
    \delta\phi = \exp(in\theta + \frac{i\theta_1}{2\pi}\theta), n\in \Z\\
    \delta\phi = i\exp(in\theta + \frac{i\theta_1}{2\pi}\theta), n\in \Z
\end{align*}
One can easily check that all the eigenfunctions satisfy the boundary condition $\delta\phi(2\pi) = {U_*\delta\phi(0) = \exp(i\theta_1)\delta\phi(0)}$.
Now we take the product of the eigenvalues of $-d^2/d\theta^2$:
\begin{equation*}
\det(-\frac{d^2}{dt^2})|_{\theta_1} = \prod_{n\in\Z}(n + \frac{\theta_1}{2\pi})^2(n + \frac{\theta_1}{2\pi})^2 = \left(\frac{\theta_1}{2\pi}\prod_{n\geq 1}(\frac{\theta_1^2}{(2\pi)^2} - n^2)\right)^4 \propto \sin(\theta_1/2)^4
\end{equation*}
where we have ignored an infinite proportionality constant. We have only computed the determinant restricted to the $\theta_1$ summand of the tangent space. To compute the full determinant we need to take product over all the summands:
\begin{equation*}
    \sqrt{\det(-\frac{d^2}{dt^2})} = \prod_i\sqrt{\det(-\frac{d^2}{dt^2})|_{\theta_i}} = \prod_i \sin(\theta_i/2)^2
\end{equation*}
For the fermionic determinant we use the basis
\begin{align*}
    \psi^{0,1} = \exp(in\theta + i\frac{\theta_1}{2\pi}\theta), n\in\Z\\
    \psi^{1,0} = \exp(in\theta + i\frac{\theta_1 + \gamma}{2\pi}\theta), n\in\Z
\end{align*}
Hence
\begin{equation*}
    \det(-\frac{d}{dt})_{0,1}\det(-\frac{d}{dt})_{1,0} = \prod_{n\in\Z, i} (n + \frac{\theta_i}{2\pi})(n + \frac{\theta_i + \gamma}{2\pi}) = \prod_i \sin(\theta_i/2)\sin((\theta_i + \gamma)/2)
\end{equation*}
So the final result for the quadratic fluctiation \eqref{quadratic_fluctuation} is
\begin{equation*}
    \prod_i \frac{\sin((\theta_i + \gamma)/2}{\sin(\theta_i/2)} = \prod_i \exp(-i\gamma/2)\frac{1 - \exp(i\theta_i)\hat{y}}{1 - \exp(i\theta_i)}
\end{equation*}
One can rewrite this expression in a different way using the character of torus action $U(1)\times U(1)\times...$ generated by $-V, -Z^1,-Z^2,...$ (note the minus sign). We decompose the tangent space at $P$ into one dimensional weight spaces of the torus. We denote the one dimensional characters as $\chi_i(t,z_i)$ where $i$ labels the one dimensional weight spaces and $(t,z_i)\in U(1)\times U(1)\times...$. Then the contribution from $P$ can be rewritten as
\begin{equation*}
    \hat{y}^{-\dim_\C \tilde{M}/2}\prod_i \frac{1 - \hat{y}\chi_i^{-1}(P)}{1 - \chi_i^{-1}(P)}
\end{equation*}
Now let us add the overall sign $(-1)^{-\dim_\C \tilde{M}/2}$.
The full superconformal index is
sum over the fixed points
\begin{equation}\label{index_abstract}
    \Tr((-1)^F\exp(-\mu\{Q, Q^\dagger\})t^{L_t}y^{L_y}z_i^{L_i}) = \sum_P (-\hat{y})^{-\dim_\C \tilde{M}/2}\prod_i \frac{1 - \hat{y}\chi_i^{-1}(P)}{1 - \chi_i^{-1}(P)}
\end{equation}

For some $M$ and $\tilde{M}$, and certainly for the instanton moduli spaces, this index equals the equivariant $\chi_y$ genus of $\tilde{M}$. To explain this concept, we recall that there is a torus action on $\tilde{M}$. The torus action naturally lifts to an action on $H^q(\Omega^p, \tilde{M})$ taken in the Zariski topology. Our convention is that if $\omega \in H^0(\Omega^p, \tilde{M})$ is a global differential form then the action is via inverse pullback $(t,z_i) \cdot\omega = (t^{-1} z_i^{-1})^*\omega$. Assuming the character $\ch$ of the torus action on $H^q(\Omega^p, \tilde{M})$ is a well-defined infinite Laurent series in $t,z_i$ then the equivariant $\chi_y$ genus is defined as
\begin{equation}\label{chi_y_genus}
    \sum_{p,q}(-1)^q (-\hat{y})^{p - \dim_\C \tilde{M}/2} \ch H^q(\Omega^p, \tilde{M}) = \sum_p (-\hat{y})^p \chi(\Omega^p)
\end{equation}
where the equivariant Euler characteristic $\chi(\Omega^p) = \sum_q (-1)^q\ch H^q(\Omega^p)$ is an alternating sum of {the characters of $H^q(\Omega^p)$}. The equivariant $\chi_y$ genus is an infinite Laurent series in $t,z_i$ and is known to resum to \eqref{index_abstract} for the resolved instanton moduli spaces \cite{Nakajima}. 
The characters are well-defined for all the examples we consider in this paper. In fact, under the following assumptions which are satisfied by the resolved instanton moduli spaces:
\begin{enumerate}
    \item M is affine
    \item The weight spaces of $H^0(\Omega^0, M)$ under $t$ action generated by $-V = -ID$ are all finite dimensional.
    \item The $t$-weights of $H^0(\Omega^0, M)$ are nonnegative and the only element with zero weight is the identity function.
    \item $\pi: \tilde{M}\to M$ is a $t,z_i$ equivariant projective morphism. Here we assume that $z_i$ also act on $M$ in a natural way.
\end{enumerate}
It can be shown that \cite{Nakajima}
\begin{enumerate}
    \item The $t$-weight spaces of $H^q(\Omega^p, \tilde{M})$ are all finite dimensional
    \item The $t$-weights of $H^q(\Omega^p, \tilde{M})$ are all bounded below
\end{enumerate}
{We briefly sketch the argument here.} First we pushforward the sheaves $\Omega^p$ on $\tilde{M}$ to obtain the higher direct image sheaves $R^q\pi_* \Omega^p$ on $M$. Since $M$ is affine, the global sections $H^0(R^q\pi_* \Omega^p, M)\cong H^q(\Omega^p, \tilde{M})$. Since the map $\pi$ is projective, $R^q\pi_* \Omega^p$ are all coherent sheaves on $M$ and hence $H^0(R^q\pi_* \Omega^p, M)$ is a finitely generated module over $H^0(\Omega^0, M)$. Hence $H^q(\Omega^p, \tilde{M})$ is also finitely generated over $H^0(\Omega^0, M)$. We can pick generators $x_1,...,x_n \in H^q(\Omega^p, \tilde{M})$ and any element of $H^q(\Omega^p, \tilde{M})$ can be obtained by acting on these $x_i$ with elements of $H^0(\Omega^0, M)$. Since $H^0(\Omega^0, M)$ has only positive $t$-weights, we deduce that the weights of $H^q(\Omega^p, \tilde{M})$ is bounded below. Since the only element of $H^0(\Omega^0,M)$ with zero $t$-weight is the identity, we see that $t$-weight spaces of $H^q(\Omega^p, \tilde{M})$ are all finite dimensional.

These results suggest that we should Taylor expand the rational function at $t = 0$ to recover the infinite Laurent series \eqref{chi_y_genus}. We define the integer valued function $\mathcal{C}(L_t,L_y, L_i)$ as the coefficients before $t^{L_t}y^{L_y}z_i^{L_i} = t^{L_t + L_y}\hat{y}^{L_y}z_i^{L_i}$ in the Taylor expansion, then it can be written as
\begin{equation*}
    \mathcal{C}(L_t,L_y, L_i) = (-1)^{L_y}\sum_{q\geq 0}(-1)^q d(q,L_t + L_y,L_i)
\end{equation*}
where $d(q,L_t+L_y.L_i)$ denotes the dimension of the $t^{L_t + L_y}z_i^{L_i}$ weight space of $H^q(\Omega^p, {\tilde{M}}),p = L_y + \dim_\C\tilde{M}/2$. In this paper we are interested in the growth of $\mathcal{C}$ when $L_t$ is large with $L_y = 0$.

Finally we are ready to prove that the index has a finite limit as $t\to 0$ predicted by the $\mathfrak{osp}(4^*|4)$ symmetry. To prove this proposition we use the geometric intepretation of this index
\begin{equation}\label{chi_y_genus_2}
    \sum_{p\geq 0}\left(-\frac{y}{t}\right)^{p - \dim_\C(\tilde{M})/2}\chi(\Omega^p)
\end{equation}
Each $\chi(\Omega^p)$ is a rational function in $t,z_i$:
\begin{equation*}
    \chi(\Omega^p) = \sum_{P} \frac{\sum_{i_1<i_2<...,<i_p}\chi_{i_1}(P)...\chi_{i_p}(P)}{\prod_i(1 - \chi_i(P))}
\end{equation*}
The sum is over all fixed points $P$ under the torus action $t,z_i$.
First we show that all $\chi(\Omega^p)$ have finite limits as $t\to 0$. There are three possible limits for each $\chi_i$ as $t\to 0$: finite, infinite or zero. In all cases both $(1 - \chi_i(P))^{-1}$ and $\chi_i(P)(1 - \chi_i(P))^{-1}$ have finite limits. Since each term is a product of $(1 - \chi_i(P))^{-1}$ and $\chi_i(P)(1 - \chi_i(P))^{-1}$ we see that $\chi(\Omega^p)$ also has a finite limit as $t\to 0$.

This result immediately implies half of the proposition: when the holomorphic degree $p\leq \dim_\C(\tilde{M})$ the term
\begin{equation*}
    \left(-\frac{y}{t}\right)^{p - \dim_\C(\tilde{M})}\chi(\Omega^p)
\end{equation*}
has a finite limit as $t\to 0$.  Now let us use the $y\to y^{-1}$ symmetry to prove that when $p > \dim_\C(\tilde{M})$ this expression also has a finite limit. The $y \to y^{-1}$ symmetry implies
\begin{equation*}
    \left(-\frac{y}{t}\right)^{p - \dim_\C(\tilde{M})/2}\chi(\Omega^p)  + \left(-\frac{y}{t}\right)^{p' - \dim_\C(\tilde{M})/2}\chi(\Omega^{p'})
\end{equation*}
is invariant under $y\to y^{-1}$ whenever $p + p' = \dim_\C M$. Hence
\begin{equation*}
    \left(-\frac{1}{t}\right)^{p - \dim_\C(M)/2}\chi(\Omega^p)  = \left(-\frac{1}{t}\right)^{p' - \dim_\C(M)/2}\chi(\Omega^{p'})
\end{equation*}
which implies
\begin{equation*}
    \chi(\Omega^{p'}) = \chi(\Omega^p)t^{p' - p}
\end{equation*}
We know $|\dim_\C(M)/2 - p'| = \frac{1}{2}|p' - p|$ and so if $p'> p$
\begin{equation*}
    \chi(\Omega^{p'})t^{\dim_\C(M)/2 - p'} = \chi(\Omega^p) t^{\dim_\C \tilde{M}/2 - p}\to 0
\end{equation*}
as $t\to 0$. Hence every term in the superconformal index \eqref{chi_y_genus_2} has a finite limit as $t\to 0$.
.
\section*{Quantum mechanics on the Hilbert scheme}
In this section we specialize to the case $\tilde{M} = {\text{Hilb}}_K(\C^2)$ of Hilbert scheme of $K$-points on $\C^2$, denoted. It resolves the singularity of $M = \Sym^K(\C^2)$ the $K$th symmetric product of $\C^2$.
The Hilbert scheme of $K$-points is described by the following ADHM data \cite{nakajima1999lectures}
\begin{eqnarray}\label{ADHM_data}
    [X, \tilde{X}] = 0,
    \left[X,X^\dagger\right] + [\tilde{X},\tilde{X}^\dagger] + QQ^\dagger - \tilde{Q}^\dagger\tilde{Q} = \xi>0
\end{eqnarray}
quotient by the action of $U(K)$ given by
\begin{equation*}
    (X, \tilde{X}, Q)\to (gXg^{-1}, g\tilde{X}g^{-1}, gQ, \tilde{Q}g^{-1}), g\in U(K)
\end{equation*}
where $X,\tilde{X}$ are $K$ by $K$ matrices and $Q\in\C^K$ is a column vector. $\tilde{Q}\in \C^K$ is a row vector. The case $\xi = 0$ corresponds to $M = \Sym^K(\C^2)$. This is a hyperkahler quotient and so ${\text{Hilb}}^K(\C^2)$ has a natural hyperKahler metric which is the restriction of the metric
\begin{equation*}
    |\delta X|^2 + |\delta\tilde{X}|^2 + |\delta Q|^2 + |\delta\tilde{Q}|^2
\end{equation*}
to the orthogonal complement to the $U(K)$ orbit in the \eqref{ADHM_data}. It is also known that this Hilbert scheme is isomorphic (via the identity map for any $\xi$) to the following noncompact (in the Euclidean topology) algebraic variety:
\begin{equation*}
    [X, \tilde{X}] = 0/{(X,\tilde{X}, Q)\sim (gXg^{-1}, g\tilde{X}g^{-1},gQ)}, g\in GL(k;\C)
\end{equation*}
subject to the stability condition: the span of $\left<X^n\tilde{X}^mQ,m,n\geq 0\right>  = \C^K$.
In this representation the complex structure of the Hilbert scheme is manifest.

The potential $C$ has the following form:
\begin{equation*}
    C = \frac{1}{2}(|X|^2 + |\tilde{X}|^2 + |Q|^2 + |\tilde{Q}|^2)
\end{equation*}
It is $U(K)$ invariant and descends to a well-defined function on ${\text{Hilb}}_K(\C^2)$ and $\Sym^K(\C^2)$. The dilation vector field $D$ on $\Sym^K(\C^2)$ is just the flat space {homothetic} vector field.

In addition to $t$, there is another global $U(1)$ isometric action called $x$ and action of $t$ and $x$ are given by:
\begin{equation*}
    (X, \tilde{X}, Q, \tilde{Q})\mapsto (t^{-1}x X, t^{-1}x^{-1} \tilde{X}, Q, t^{-2}\tilde{Q})
\end{equation*}
To make our calculation easier we introduce the notation $(q_1,q_2)\in U(1)^2$:
\begin{equation*}
    q_1 = tx, q_2 = t/x
\end{equation*}
The corresponding quantum numbers are naturally interpreted as the two angular momenta of the supersymmetric particle. 

To compute the superconformal index \eqref{index_abstract} we need to know the fixed points of the $U(1)^2$ action and the {characters} at the fixed points. The fixed points are parametrized by young tableaux with $K$ boxes \cite{nakajima1999lectures}. The appendix summarizes our convention for young tableaux. The character at a fixed point $Y$ (a young tableaux) is given by:
\begin{equation*}
    \sum_{s\in Y} q_1^{-L(s) - 1}q_2^{A(s)} + q_1^{L(s)}q_2^{-A(s) - 1}
\end{equation*}
Substitute in the localisation formula \eqref{index_abstract} the superconformal index is (up to an overall sign $(-1)^{\dim_\C\tilde{M}/2}$)
\begin{equation}\label{index}
    Z_K(q_1, q_2, \hat{y}) = \sum_{|Y| = K}\frac{(1 - \hat{y}q_1^{L(s) + 1}q_2^{-A(s)})(1 - \hat{y}^{-1}q_1^{L(s)}q_2^{-A(s) - 1})}{(1 - q_1^{L(s) + 1}q_2^{-A(s)})(1 - q_1^{L(s)}q_2^{-A(s) - 1})}
\end{equation}
And the function $\mathcal{C}$ (again up to an overall {sign} which will not be important since we are only interested in $|\mathcal{C}|$) is the Taylor coefficients of $Z_K$ written in variables $t,x,y$:
\begin{equation*}
    Z_K(t,x,y) = \sum_{L_t\geq 0,  L_y, L_x}\mathcal{C}(L_t,L_y,L_x,K)t^{L_t}y^{L_y}x^{L_x}
\end{equation*}
In this paper we investigate the asymptotic of $\mathcal{C}(L, 0, 0, K)$ in the Cardy limit $L, K\to \infty$ with $L/K$ fixed. The main tool we will use is the following generating function:
\begin{equation*}
    Z(q_\tau, q_1,q_2,\hat{y}) = \sum_{K\geq 0}q_\tau^K Z_K(q_1, q_2, \hat{y})
\end{equation*}
The numbers $\mathcal{C}(L_t,L_x,L_y,K)$ are the Taylor coefficients of the expansion
\begin{equation*}
    Z(q_\tau, q_1,q_2,\hat{y}) = \sum_{L_t\geq 0, L_y,L_x, K\geq 0}\mathcal{C}(L_t, L_x,L_y,K)t^{L_t}x^{L_x}y^{L_y}q_\tau^K
\end{equation*}
The generating function admits the following resummation \cite{Rains} which will be essential to our proof:
\begin{equation*}
    Z(q_\tau,q_1,q_2,\hat{y}) = \exp\left(\frac{1}{n}\frac{q_\tau^n}{1-q_\tau^n}\frac{\hat{y}^{-n}(1 - \hat{y}^nq_1^n)(1 - \hat{y}^nq_2^n)}{(1-q_1^n)(1 - q_2^n)}\right)
\end{equation*}
The infinite sum inside the exponential converges to a holomoprhic function in the following domain:
\begin{eqnarray}\label{domain}
    |q_\tau|<1, |q_1|<1, |q_2|<1, |\Re(\hat{m})|< \frac{3}{4}\Re(\beta), \hat{y} = \exp(-\hat{m}), q_\tau = \exp(-\beta)
\end{eqnarray}
It can actually be analytically continued to a large domain but throughout this paper we will only work with this domain. The factor $3/4$ is chosen because later in our proof $\Re (\hat{m})$ cannot be too close to $\Re (\beta)$. We will extract the Taylor coefficients $\mathcal{C}$ using {a contour integral}
\begin{equation*}
    \mathcal{C}(2L, 0,0, K) = \frac{1}{(2\pi i)^4}\int \frac{dq_\tau}{q_\tau^{K+1}}\frac{dq_1}{q_1^{L+1}} \frac{dq_2}{q_2^{L+1}}\frac{d\hat{y}}{\hat{y}} \exp\left(\frac{1}{n}\frac{q_\tau^n}{1-q_\tau^n}\frac{\hat{y}^{-n}(1 - \hat{y}^nq_1^n)(1 - \hat{y}^nq_2^n)}{(1-q_1^n)(1 - q_2^n)}\right)
\end{equation*}
{Then we will apply the saddle point method to compute its asymptotic}. Notice that we have replaced $L$ with $2L$ because when the power of $x$ is zero, the power of $t$ is always even.
{The saddle point method predicts that in the limit $L,K>>1$, the integral of $\mathcal{C}$ is dominated by a saddle point near $q_\tau = q_1 = q_2 = 1, \hat{y} = -1$ and the leading order growth of $\mathcal{C}$ is the value of the integrand at the saddle point.}
The prediction is:
\begin{equation}\label{imprecise_prediction}
    \mathcal{C}(2L,0,0,K)\sim \exp(2\sqrt{2} \pi \sqrt{L}K^{1/4}24^{-1/4})\approx \exp(4.0146\sqrt{L}K^{1/4})
\end{equation}
as $K,L\to\infty$ when $K/L$ fixed. {A nonrigorous derivation of this asymptotic will appear in the next section. In this paper we will prove a more precise version of the asymptotic}
\begin{eqnarray}\label{asymptotic}
    \mathcal{C}(2L,0,0,K) = \exp(2\sqrt{2}\pi 24^{-1/4}\sqrt{L}K^{1/4} + P)(\cos(2\sqrt{2}\pi 24^{-1/4}\sqrt{L}K^{1/4} + Q) + o(1))
\end{eqnarray}
where $P,Q$ are real polynomials in $K,\log(K)$ for fixed $K/L$ and satisfy the following bound
\begin{equation*}
    P = \mathcal{O}_{K/L}(K^{2/4}), Q= \mathcal{O}_{K/L}(K^{2/4})
\end{equation*}
The $o(1)$ term goes to zero as $K\to \infty$.
We will prove \eqref{asymptotic} under the assumption $L/K$ is fixed and at least 2. Notice that there is an oscillating {cosine} factor. As a result, $\mathcal{C}$ can be both positive and negative which is confirmed by numerical data. We are unable to bound the cosine from below but we believe that for generic choice of $K/L$ and for sufficiently many $K$ the cosine should not be too small and hence the exponential term should dominate the growth. 

\section*{A nonrigorous derivation of the asymptotic}
In this section we present a nonrigorous derivation of the asymptotic based on the saddle point method originally used by Hardy and Ramanujan \cite{HardyRamanujan} in their study of the partition functions. Recall that we have expressed our coefficients $\mathcal{C}$ as a contour integral.
\begin{equation*}
     \mathcal{C}(2L, 0,0, K) = \frac{1}{(2\pi i)^4}\int \frac{dq_\tau}{q_\tau^{K+1}}\frac{dq_1}{q_1^{L+1}} \frac{dq_2}{q_2^{L+1}}\frac{d\hat{y}}{\hat{y}} Z[q_\tau,q_1,q_2, \hat{y}]
\end{equation*}
where the contours for $q_1,q_2,q_\tau$ are circles centred at the origin with {radii} less than one. The contour for $\hat{y}$ is the unit circle. All contours are traversed counterclockwise. To make our computation easier, we perform a change of variables
\begin{equation}\label{change_of_variables}
    q_\tau = \exp(-\beta), q_1 = \exp(-\epsilon_1), q_2 = \exp(-\epsilon_2), \hat{y} = \exp(-\hat{m})
\end{equation}
{and} rewrite the contour integral as
\begin{equation}\label{contour_representation}
    \mathcal{C}(2L, 0,0, K) = \frac{1}{(2\pi i)^4}\int d\epsilon_1d\epsilon_2d\hat{m} d\beta \exp(\log(Z) + K\beta + L\epsilon_1 + L\epsilon_2)
\end{equation}
{Hereafter} $\log(Z)$ always means
\begin{equation*}
    \log(Z) = \sum_{n\geq 1}\frac{1}{n}\frac{q_\tau^n}{1-q_\tau^n}\frac{\hat{y}^{-n}(1 - \hat{y}^{n}q_1^n)(1 - \hat{y}^{n}q_2^n)}{(1-q_1^n)(1 - q_2^n)}
\end{equation*}
The contours for $\epsilon_1,\epsilon_2,\beta$ are straight vertical lines with positive real parts. We will specify the real parts later. The imaginary parts go from $-i\pi$ to $i\pi$. The contour for $\hat{m}$ is $[0, 2\pi i]$. Hereafter closed intervals represent straight lines on the complex plane.

The saddle point method tells us that the large $K$ asymptotic of the function $\mathcal{C}(2L,0,0,K)$ is dominated by the integral around the maximum of the integrand $\exp(\log(Z) + K\beta + L\epsilon_1 + L\epsilon_2)$. A saddle point is a point at which the first derivatives of the integrand vanish. We do not know a priori where the maximum is attained and so we make a guess: we assume that the integrand is maximized near $q_1 = q_2 = q_\tau = 1$. For general $q_1,q_2,q_\tau, \hat{y}$ the function $\log(Z)$ is complicated. However, when $q_1 \approx 1, q_2\approx 1, q_\tau\approx 1$ we can make {the following approximations}:
\begin{eqnarray*}
    1 - \hat{y}^nq_1^n\approx 1 - \hat{y}^n, 1 - \hat{y}^nq_2^n\approx 1 - \hat{y}^n\\ 1-q_\tau^n\approx n\beta, 1 - q_1^n\approx n\epsilon_1, 1 - q_2^n\approx n\epsilon_2
\end{eqnarray*}
As a result we can approximate\footnote{Identities involving polylogarithms are summarized in the appendix}
\begin{equation}\label{leading_asymptotic}
    \log(Z)\approx \sum_{n\geq 1}\frac{1}{n}\frac{\hat{y}^{-n}(1- \hat{y}^n)^2}{n^3\epsilon_1\epsilon_2\beta} = -\frac{\hat{m}^2(\hat{m} - 2\pi i)^2}{24\epsilon_1\epsilon_2\beta}
\end{equation}
And so we are looking for the maximum of
\begin{equation}\label{leading_integrand}
    \exp\left(-\frac{\hat{m}^2(\hat{m}-2 \pi i)^2}{24\epsilon_1\epsilon_2\beta} + K\beta + L\epsilon_1 + L\epsilon_2\right)
\end{equation}
We take partial derivatives with respect to $\epsilon_1,\epsilon_2,\beta,\hat{m}$ and set all the partial derivatives to be zero. The saddle point equations are:
\begin{eqnarray*}
    \frac{\hat{m}^2(\hat{m}-2\pi i)^2}{24\epsilon_1^2\epsilon_2\beta} + L = 0\\
    \frac{\hat{m}^2(\hat{m}-2\pi i)^2}{24\epsilon_1\epsilon_2^2\beta} + L = 0\\
    \frac{\hat{m}^2(\hat{m}-2\pi i)^2}{24\epsilon_1\epsilon_2\beta^2} + K = 0\\
    \partial_{\hat{m}} \frac{\hat{m}^2(\hat{m}-2 \pi i)^2}{24\epsilon_1\epsilon_2\beta} = 0
\end{eqnarray*}
There are two solutions and the phases of $\epsilon_1,\epsilon_2,\beta$ are $\pi/4$ in one of the solutions and $-\pi/4$ in the other solution. In other words,
\begin{equation}\label{saddle}
    \epsilon_1 = \Re(\epsilon_1) \pm i\Re(\epsilon_1), \epsilon_2 = \Re(\epsilon_2) \pm i\Re(\epsilon_2), \beta = \Re(\beta) \pm i\Re(\beta), \hat{m} = i\pi
\end{equation}
where the real parts are given by:
\begin{eqnarray}\label{real_parts}
    \Re(\beta) = \pi\sqrt{L}K^{-3/4}24^{-1/4}2^{-1/2}\\ \Re(\epsilon_1) =\Re(\epsilon_2)=\frac{K}{L}\Re(\beta) = \pi K^{1/4}L^{-1/2}24^{-1/4}2^{-1/2},
\end{eqnarray}
We see that the saddle point values of $\epsilon_1,\epsilon_2,\beta\sim K^{-1/4}\to 0$ as $K\to \infty$. Now we can deform our contours to pass through these saddle points. In other words, we take the integration contours for $\epsilon_1,\epsilon_2,\beta, \hat{m}$ to be the following straight vertical lines:
\begin{eqnarray*}
    \epsilon_1\in [\Re(\epsilon_1) - i\pi, \Re(\epsilon_1) + i\pi]\\
     \epsilon_2\in [\Re(\epsilon_2) - i\pi, \Re(\epsilon_2) + i\pi]\\
     \beta\in [\Re(\beta) - i\pi, \Re(\beta) + i\pi]\\
     \hat{m}\in[0,2\pi i]
\end{eqnarray*}
where the real parts are given in \eqref{real_parts}.
Now we just need to replace $\epsilon_1,\epsilon_2,\beta,\hat{m}$ in \eqref{leading_integrand} by \eqref{saddle} to obtain the desired asymptotic \eqref{imprecise_prediction}.

In order to make this calculation rigorous, we need to answer the following questions:
\begin{enumerate}
    \item Is there any other saddle point away from $q_1,q_2,q_\tau\approx 1$ which contribute to the leading asymptotic?
    \item What is the error in the approximation \eqref{leading_asymptotic} and does it affect the leading asymptotic at the saddle point?
    \item The first derivative of $\log(Z) + K\beta + L \epsilon_1 + L\epsilon_2$ at the saddle point\eqref{saddle} is not zero due to higher correction terms. If we expand the integrand around the saddles we get linear fluctuations in addition to quadratic fluctuations. Are these linear fluctuations around the saddle points significant?
\end{enumerate}
In this paper we will see that the answers to first two questions are both no. The answer to the last question is yes and we need to deform the contours to pass through more accurate saddle points.

\section*{Outline of the Proof}
Now we outline the proof of the asymptotic \eqref{asymptotic} under the assumption $L/K\geq 2$. There are several steps:
\begin{enumerate}
    \item Step 1: We bound the growth away from the two saddle points.
    We divide the range of integration into
\begin{equation}\label{saddle_region}
    |\Im(\epsilon_1)|<100\Re(\epsilon_1), |\Im(\epsilon_2)|<100\Re(\epsilon_2), |\Im(\beta)|<100\Re(\beta)
\end{equation}
which contains the two saddle points and its complement. In this region, the imaginary parts {$\Im(\epsilon_1),\Im(\epsilon_2),\Im(\beta)$} are comparable to the real parts { $\Re(\epsilon_1),\Re(\epsilon_2),\Re(\beta)$} and we hope to approximate $\log(Z)$ by its leading asymptotic \eqref{leading_asymptotic}. In the complement we will show that the integral is bounded by $\mathcal{O}(\exp(3.38K^{1/4}L^{1/2}))$ which decays exponentially relative to the predicted leading order growth \eqref{asymptotic}. Notice that the range of $\hat{m}$ is still $[0, 2\pi i]$. Once we prove this bound, the integral over the complement of \eqref{saddle_region} is absorbed into the $o(1)$ term in \eqref{asymptotic}. Hence we only need to study the integral over \eqref{saddle_region}. And in this region the same method allow{s} us to discard the integral when $\hat{m}\in [0, i\pi/4]$ or $[7i\pi/4,2\pi i]$ by bounding the integral in that region by $\mathcal{O}(\exp(3.93K^{1/4}L^{1/2}))$. This is as far as we can prove using this method. In particular we are not able to restrict the contour of $\hat{m}$ to an arbitrarily small neighborhood of $i\pi$ at this stage. We will be able to do so after we compute an asymptotic for $\log(Z)$ in the saddle region.

\item Step 2: In the region
\begin{equation}\label{saddle_region_2}
    |\Im(\epsilon_1)|<100\Re(\epsilon_1), |\Im(\epsilon_2)|<100\Re(\epsilon_2), |\Im(\beta)|<100\Re(\beta), \hat{m}\in [i\pi/4, 7i\pi/4]
\end{equation}
We will compute the asymptotic of the integrand and {prove} the following
\begin{eqnarray}\label{saddle_region_asymptotic}
    \log(Z) = 
    \underbrace{-\frac{\hat{m}^2(\hat{m}-2\pi i)^2}{24\beta\epsilon_1\epsilon_2}}_{\sim K^{3/4}} + \underbrace{\frac{f_1(\hat{m})}{\beta\epsilon_1} + \frac{f_2(\hat{m})}{\beta\epsilon_2} + \frac{f_3(\hat{m})}{\epsilon_1\epsilon_2}}_{\sim K^{2/4}} + \underbrace{E(\beta, \epsilon_1,\epsilon_2,\hat{m})}_{O_{K/L}(K^{1/4})}
\end{eqnarray}
for some holomorphic functions $f_1(\hat{m}), f_2(\hat{m}), f_3(\hat{m})$ \eqref{f_i_expression} chosen in a way so that $E$ is the second subleading correction.
We will bound the error term $E(\beta, \epsilon_1,\epsilon_2,\hat{m})$ and its derivative $\nabla E$. The bound on $|E|$ allows us to show that the leading order asymptotic does indeed {come} from the $-\hat{m}^2(\hat{m}-2\pi i)^2/(24\beta\epsilon_1\epsilon_2)$ term.
We will show that $|\nabla E|$ is small so that the first derivatives of $\log(Z)$ can be well approximated by the first derivatives of the three terms $f_1/(\beta\epsilon_1), f_2/(\beta\epsilon_2),f_3/(\epsilon_1\epsilon_2)$.

Here are the estimates:
\begin{eqnarray}\label{bound_E}
    |E| = \mathcal{O}_{K/L}(K^{1/4}), \partial_{\hat{m}} E = \mathcal{O}_{K/L}(K^{1/4}\log K), \partial_{\epsilon_1,\epsilon_2,\beta}E = \mathcal{O}_{K/L}(K^{2/4})
\end{eqnarray}
The notation $\partial_{\epsilon_1,\epsilon_2,\beta}A = \mathcal{O}(B)$ means $\partial_\beta A = \mathcal{O}(B), \partial_{\epsilon_1}A = \mathcal{B}, \partial_{\epsilon_2}A = \mathcal{O}(B)$. In other words, all three derivatives satisfy the same bound $\mathcal{O}(B)$.
Inuitively, $E$ should be well approximated by a linear sum of $1/\epsilon_1,1/\epsilon_2,1/\beta$ with coefficients depending on $\hat{m}$. Hence differentiating with respect to any one of $\epsilon_1,\epsilon_2,\beta$ increases the growth of $E$ by $K^{1/4}$. However, differentiating with respect to $\hat{m}$ does not change the growth of $E$ by any polynomial power $K$ (so $\log K$ is allowed). We will also need to show that this bound holds when $\hat{m}$ has a small real part or when the real parts of $\epsilon_1,\epsilon_2,\beta$ undergo a small perturbation. 

\item Step 3:
Now we study the leading asymptotic inside the exponential
\begin{equation}\label{leading_exponent}
    -\frac{\hat{m}^2(\hat{m}-2\pi i)^2}{24\beta\epsilon_1\epsilon_2} + K\beta + L\epsilon_1 + L\epsilon_2
\end{equation}
and find the maximum of its real part in the region \eqref{saddle_region}. Since we will eventually take the exponential of this expression, the real part controls the modulus of the exponential.
We show that the two saddle points are precisely the location at which the real parts are maximized. We also compute the Hessian of the real part at these two saddle points and show that they are both negative definite. To simplify the notation we perform a change of variable
\begin{equation*}
    \epsilon_1 = \Re(\epsilon_1) \sigma_1, \epsilon_2 = \Re(\epsilon_2)\sigma_2,\beta = \Re(\beta) \sigma_3, \hat{m} = \sigma_4
\end{equation*}
so that $\sigma_i$ are all order 1 variable. The contours for $\sigma_1,\sigma_2,\sigma_3$ all have real part equal to 1 and imaginary part in $[-100,100]$. We temporarily allow the contour of $\hat{m} = \sigma_4$ to be $[0,2\pi i]$ to make the calculation simpler. If we can find the maximum over a larger region and the maximum lies in the smaller region we are interested in, then we know the maximum over the smaller region. The two approximate saddle points $\sigma^\pm$ are
\begin{equation}\label{approximate_saddle}
    \sigma_1 = \sigma_2 = \sigma_3 = 1\pm i, \sigma_4 = i\pi
\end{equation}
The leading asymptotic has vanishing first derivative at \eqref{approximate_saddle}. However, once we take into account subleading correction{s}, the derivative of $\log(Z) + K\beta + L\epsilon_1 + L\epsilon_2$ is nonzero at \eqref{approximate_saddle}. As a result, if one tries to perform {Taylor expansions} around the saddle point{s} and {compute} the Gaussian fluctuation, one gets an additional linear term in the integral which leads to difficulties. As a result we will deform the contour to pass through a pair of more accurate saddle points $\tilde{\sigma}^\pm$. The linear terms at the new saddles points are negligible.

\item Step 4: In this step we compute {the} more accurate saddle points $\tilde{\sigma}^\pm$ using the leading {term and the first subleading correction}:
\begin{equation*}
    -\frac{\hat{m}^2(\hat{m}- 2\pi i)^2}{24\beta\epsilon_1\epsilon_2} + \frac{f_3(\hat{m})}{\epsilon_1\epsilon_2} + \frac{f_1(\hat{m})}{\beta\epsilon_1} + \frac{f_2(\hat{m})}{\beta\epsilon_2} + K\beta + L \epsilon_1 + L\epsilon_2
\end{equation*}
The subleading correction is small compared with {the leading term in the limit when} $|\epsilon_1|,|\epsilon_2|,|\beta|<<1$ and therefore one can apply the implicit function theorem to show the existence of the {saddle points} to this function. The result is that leading order shift of the saddle point value of $\hat{m}$ depends on $\epsilon_1,\epsilon_2$ only:
\begin{equation*}
    \hat{m} = i\pi - (\epsilon_1 + \epsilon_2)/2 = i\pi - \epsilon_1
\end{equation*}
In the second equality we used {$\epsilon_1 = \epsilon_2$ at the saddle point}. The saddle point of $\hat{m}$ needs to lie within the domain of definition of $\hat{m}$ \eqref{domain}. As a result, we require $\Re(\tau)\geq \Re(\epsilon_1)/2$ and so $L/K\geq 2$. We denote the two new saddle points as $\tilde{\sigma}^\pm$. They are complex conjugate to each other $\overline{\tilde{\sigma}^+} = \tilde{\sigma}^-$. For $\hat{m}$ complex conjugation {is} accompanied by an addition of $2\pi i$.

\item Step 5:
We deform the contours for $\sigma_1,\sigma_2,\sigma_3,\sigma_4$ so that they pass through $\tilde{\sigma}^\pm$. The first derivative at $\tilde{\sigma}^\pm$ are now small enough so we can safely perform Taylor expansion around the saddle points and ignore the linear term{s}. We will focus on the contribution from small balls around the two saddle points and bound the error{s} away from the two {balls}. Inside the two {balls} we replace the integrand by its Taylor expansions to second order around the saddle points $\tilde{\sigma}^\pm$. As long as the radii of the two small balls tend to zero when $K\to\infty$, the Taylor expansions are guaranteed to be good approximations inside the two balls. Outside the two small balls, {the integrands are bounded by their values on the boundaries of the two balls}. The negative definiteness of the {two Hessians}plays an essential role in this bound.

\item Step 6:
In the last step, we compute the saddle point value of the integrand and show that $(\log(Z) + K \beta + L\epsilon_1 + L\epsilon_2)(\tilde{\sigma}^\pm)$ are indeed polynomials in $K, \log(K)$. The technique used in this step comes from the paper \cite{Newman}.
\end{enumerate}

\section*{Step 1: Error away from the saddle point}
In this section we begin our proof of the asymptotic \eqref{asymptotic}. First we bound the error in the region away from the saddle points:
\begin{equation}\label{saddle_complement}
    \{|\Im(\epsilon_1)|> 100|\Re(\epsilon_1)|\} \cup \{|\Im(\epsilon_2)|> 100|\Re(\epsilon_2)|\} \cup \{|\Im(\beta)|>100\Re(\beta)\}
\end{equation}
The number 100 can be replaced by any other positive integer provided it is sufficiently large. The idea of the proof already apppeared in \cite{Newman}: when we compute the asymptotic of the infinite series with $\hat{y} = -1$ and {$q_1,q_2 = 1$} in the numerator:
\begin{equation*}
    \log(Z)\approx \sum_{n\geq 1} \frac{1}{n^4}\frac{(-1)^n(1 - (-1)^n)^2}{\beta\epsilon_1\epsilon_2}\approx \frac{-4.05871}{\beta\epsilon_1\epsilon_2}
\end{equation*}
We notice that the $n=1$ term already contributes $-4(\beta\epsilon_1\epsilon_2)^{-1}$. Due to the $1/n^4$ decay the sum from $n=2...\infty$ contributes only a tiny amount to the final asymptotic. As a result we will estimate the term $n=1$ and the terms $n\geq 2$ separately. We will not place a strong estimate on the terms $n\geq 2$ because their contribution is very small. However we will place a strong estimate on the $n=1$ term because it dominates the asymptotic.

For $n\geq 2$ we bound the numerator using $|\hat{y}^{-n}(1 - \hat{y}^{n}q_1^n)(1 - \hat{y}^{n}q_2^n)|<4$ since $|{\hat{y}}| = 1, |q_1|<1 ,|q_2|<1$. This bound may seem weak but it sufficies for our purpose. For the denominator we use the following {inequality}
\begin{equation*}
    \left|\frac{1}{1 - \exp(-n\Re(\beta) - i n\Im(\beta))}\right| \leq \left|\frac{1}{1 - \exp(-n\Re(\beta))}\right|
\end{equation*}
which is proved by drawing a circle centred at the origin with radius $\exp(-n\Re(\beta))$ in the complex plane and minimize the distance between the circle and the point $1$. One can write down similar estimates with $\beta$ replaced by $\epsilon_1,\epsilon_2$. Hence we obtain the following:
\begin{eqnarray}\label{ngeq2bound_1}
    \left|\Re\left(\sum_{n\geq 2}-\frac{1}{n}\frac{q_\tau^n}{1-q_\tau^n}\frac{\hat{y}^{-n}(1 - \hat{y}^{n}q_1^n)(1 - \hat{y}^{n}q_2^n)}{(1-q_1^n)(1 - q_2^n)}+ L \epsilon_1+ L\epsilon_2+K\beta\right)\right|\\\leq \sum_{n\geq 2}\frac{1}{n}\frac{4\exp(-n\Re(\beta))}{(1 - \exp(-n\Re(\beta)))(1 - \exp(-n\Re(\epsilon_1)))^2} + L\Re(\epsilon_1) + L\Re(\epsilon_2) + K\Re(\beta)
\end{eqnarray}
Now we still need to bound the first term on the right hand side. To simplify the notation{,} let us write $x = n\Re(\beta)>0$. The function
\begin{equation}\label{auxiliary_estimate}
    \frac{\exp(-x)}{(1 - \exp(-x))(1- \exp(-x K/L))^2} = \frac{L^2}{K^2x^3} + \mathcal{O}_{K/L}(x^{-2})
\end{equation}
as $x\to 0$. This $\mathcal{O}_{K/L}(x^{-2})$ bound is valid in a neighborhood of $0$. We want to show that it is valid everywhere. Since the function decays exponentially as $x\to \infty$ and $x^{-3}$ decay faster than $x^{-2}$ we see that the bound \eqref{auxiliary_estimate} is valid for all $x>0$.
Hence \eqref{ngeq2bound_1} is bounded by
\begin{equation*}
\sum_{n\geq 2}\frac{4 L^2}{n^4\Re(\beta)^3K^2} + \mathcal{O}_{K/L}(\Re(\beta)^{-2}) + K\Re(\beta)+L\Re(\epsilon_1) + L\Re(\epsilon_2)
\end{equation*}

Next let us bound the term $n=1$. When $|\Im(\epsilon_1)|\geq 100 \Re(\epsilon_1)$ we have
\begin{equation*}
    |1- q_1| \geq |\Im(1 - q_1)| = |\exp(-\Re(\epsilon_1))\sin(\Im(\epsilon_1))|\geq \frac{99}{100}\Im(\epsilon_1)\geq 99\Re(\epsilon_1)
\end{equation*}
when $K$ is sufficiently large. Hence in the region $\Im(\beta)\geq 100\Re(\beta)$ we have the following bound on the $n=1$ term
\begin{equation}\label{n=1bound}
    \left|\frac{q_\tau \hat{y}^{-1}(1 - \hat{y}q_1)(1 - \hat{y}q_2)}{(1 - q_\tau)(1 - q_1)(1 - q_2)}\right|\leq \frac{4}{99\Re(\beta)\Re(\epsilon_2)\Re(\epsilon_1)}
\end{equation}
Again {the numerator is bounded by $4$}: $|q_\tau \hat{y}^{-1}(1 - \hat{y}q_1)(1 - \hat{y}q_2)| < 4$. The same bound \eqref{n=1bound} remains valid in the region $|\Im(\epsilon_2)|\geq 100 \Re(\epsilon_2)$ and $|\Im(\beta)|\geq 100 \Re(\beta)$ by the same argument. Hence away from the saddle point \eqref{saddle_complement} the real part of the expression inside the expontial $\Re(\log(Z) + K\beta + L\epsilon_1 + L\epsilon_2)$ is bounded by the following
\begin{equation*}
    \frac{4}{99\Re(\beta)\Re(\epsilon_2)\Re(\epsilon_1)} + \sum_{n\geq 2}\frac{4 L^2}{n^4\Re(\beta)^3K^2} + \mathcal{O}_{K/L}(\Re(\beta)^{-2})+ K\Re(\beta) + L\Re(\epsilon_1)+L\Re(\epsilon_2)
\end{equation*}
Since $\Re(\beta)^{-2}$ grows slower than $1/(\Re(\beta)\Re(\epsilon_2)\Re(\epsilon_1))$. We can ignore this term after replacing $99$ by $98$ in the first term. We also {substitute in} $\Re(\epsilon_1) = \Re(\epsilon_2) = K\Re(\beta)/L$ {and bound the expression above by}
\begin{equation*}
    (\frac{4}{98} + \sum_{n\geq 2}\frac{4}{n^4})\frac{1}{\Re(\beta)\Re(\epsilon_1)\Re(\epsilon_2)} + K\Re(\beta) + L\Re(\epsilon_1)+L\Re(\epsilon_2)
\end{equation*}
Now we replace $\Re(\beta), \Re(\epsilon_1), \Re(\epsilon_2)$ by \eqref{saddle} and the expression becomes
\begin{equation*}
 3.37...\sqrt{L}K^{1/4}< 3.38 \sqrt{L}K^{1/4} 
\end{equation*}
And so the integral over this region is bounded by $\exp(3.38\sqrt{L}K^{1/4})$ times a positive constant (the volume of the integration region and factors of $2\pi i$). {The integral over this region decays exponentially relative to the predicted growth $\exp(4.0146\sqrt{L}K^{1/4})$} and contributes to the o(1) factor of the asymptotic\eqref{asymptotic}.

Hence from now on our region of integration will be \eqref{saddle_region}. Notice that the contour for $\hat{m}$ is still $[0,2\pi i]$. The functions $f_1,f_2,f_3$ in \eqref{saddle_region_asymptotic} develop singularities at $\hat{m} = 0$ { and could be problematic}. Hence in this section we will restrict the contour for $\hat{m}$ from $[0,2\pi i]$ to $[i\pi/4, 7i\pi/4]$. We follow exactly the same method, separating the {sum} over $n$ into $n=1$ and $n\geq 2$. The sum from $n\geq 2$ is still bounded by (up to a subleading correction $\mathcal{O}_{K/L}(K^{2/4})$ which can be neglected in the large $K$ limit)
\begin{equation*}
    \sum_{n\geq 2}\frac{4}{n^4}\frac{1}{\Re(\beta)\Re(\epsilon_1)\Re(\epsilon_2)}
\end{equation*}
However the estimates for $n=1$ is different. We need the fact that $q_1,q_2,q_\tau\approx 1$. The estimate for the denominator is
\begin{equation*}
    \frac{1}{(1- q_\tau)(1 - q_1)(1-q_2)}\leq \frac{1.01}{\Re(\beta)\Re(\epsilon_1)\Re(\epsilon_2)}
\end{equation*}
To estimate the numerator we notice that it converges to $\hat{y}^{-1}(1 - \hat{y})^2$. The function $|(1- \hat{y})^2|$ when $\hat{m}\in [0,i\pi/4]$ or $[7i\pi/4, 2\pi i]$ is bounded by $|(1 - \exp(i\pi/4))|^2 \approx 0.58$. Therefore, when $K$ is large
\begin{equation*}
    |q_\tau \hat{y}^{-1}(1 - \hat{y} q_1)(1 - \hat{y} q_2)|< 0.59
\end{equation*}
Hence in the region
\begin{equation*}
    |\Im(\beta)|\leq 100 |\Re(\beta)|,|\Im(\epsilon_2)|\leq 100 |\Re(\epsilon_2)|,|\Im(\epsilon_1)|\leq 100 |\Re(\epsilon_1)|,\hat{m}\in[0,i\pi/4]\cup [7i\pi/4,2\pi i]
\end{equation*}
when $K$ is sufficiently large, $|\Re(\log(Z)) + K \Re(\beta) + L\Re( \epsilon_1) + L\Re( \epsilon_2)|$ is bounded by
\begin{equation*}
    (0.59 \times 1.01 + \sum_{n\geq 2}\frac{4}{n^4})\frac{1}{\Re(\beta)\Re(\epsilon_1)\Re(\epsilon_2)} + K\Re(\beta) + L\Re(\epsilon_1)+L\Re(\epsilon_2)\approx 3.92 K^{1/4}\sqrt{L}
\end{equation*}
which also decays exponentially relative to the predicted growth $\exp(4.01\sqrt{L}K^{1/4})$. Hence we can restrict the contour for $\hat{m}$ to $[i\pi/4, 7i\pi/4]$ from now on.

\section*{Step 2: Asymptotic of the integrand near saddle point}
In the previous section we have shown that the integral away from the saddle point region is negligible. So it remains to compute the integral in the saddle point region. The function $\log(Z)$ is complicated. However near the saddle point $\log(Z)$ can be approximated by some simple rational function \eqref{saddle_region_asymptotic} in $\epsilon_1,\epsilon_2,\beta$. In this section we show {that} this approximation is valid for suitable choice of $f_1,f_2,f_3$ and the error term $E$ is indeed subleading. In other words, we will prove \eqref{bound_E}.
We write $\log(Z)$ as the sum of four terms and compute error estimates for each one of them:
\begin{equation}\label{four_terms}
    \log(Z) = \sum_{n\geq 1} \frac{1}{n}\frac{q_\tau^n \hat{y}^{-n} + q_\tau^n\hat{y}^nq_1^nq_2^n - q_\tau^nq_1^n - q_\tau^nq_2^n}{(1 - q_\tau^n)(1 - q_1^n)(1 - q_2^n)}
\end{equation}
Now the numerator is a sum of four terms so $\log(Z)$ is also a sum of four term.
We will compute the asymptotic of the first term near $q_1 = q_2 = q_\tau = 1$ as an example:
\begin{equation*}
    \sum_{n\geq 1}\frac{1}{n}\frac{q_\tau^n\hat{y}^{-n}}{(1 - q_1^n)(1 - q_2^n)(1 - q_\tau^n)}
\end{equation*}
We do not make any approximation of the numerator because $q_\tau^n$ creates the decay in $n$ when $n$ is large. If we perform any Taylor expansion of $q_\tau^n$ we would lose this decay and our estimates would not be valid due to the divergence of the series. Therefore we keep the numerator as it is and approximate the denominator. The function $(1 - \exp(-\beta))^{-1}$ is holomorphic in the annulus $0<|\beta|<2\pi$ and hence can be expanded in a Laurent series in this annulus
\begin{equation*}
    \frac{1}{1 - \exp(-\beta)} = \frac{1}{\beta} + \frac{1}{2} + ...
\end{equation*}
where $...$ denotes higher order terms in $\beta$. Now we do the same for $(1 - \exp(-\epsilon_1))^{-1}$ and $(1 - \exp(-\epsilon_2))^{-1}$ and multiply the three infinite Laurent series together.
\begin{equation*}
    \frac{1}{(1-q_\tau)(1 - q_1)(1 - q_2)} = \frac{1}{\beta\epsilon_1\epsilon_2} + \frac{1}{2\epsilon_1\beta} + \frac{1}{2\epsilon_2\beta} + \frac{1}{2\epsilon_1\epsilon_2} + f(\epsilon_1,\epsilon_2,\beta)
\end{equation*}
The function $f$ is an infinite series and total degree of each monomial in $f$ is at most one. Now we use the fact that the ratio $|\epsilon_1/\epsilon_2|,|\epsilon_1/\tau|$ are both bounded above and below by nonzero constants depending on $L/K$. As a result the function $f$ is bounded by
\begin{equation}\label{bound_f}
    |f| = \mathcal{O}_{K/L}(\epsilon_1^{-1}), |\nabla f| =\mathcal{O}_{K/L}(\epsilon_1^{-2})
\end{equation}
whenever $|\epsilon_1| < 1, |\epsilon_2| < 1, |\epsilon_3|<1$. The number $1$ can be replaced by any other number less than $2\pi$.
When $n|\epsilon_1| < 1, n|\epsilon_2| < 1, n|\beta| < 1$ we can write
\begin{eqnarray}\label{asymptotic_first_term}
    \frac{1}{n}\frac{q_\tau^n\hat{y}^{-n}}{(1-q_\tau^n)(1 - q_1^n)(1 - q_2^n)} = \frac{q_\tau^n\hat{y}^{-n}}{n^4\beta\epsilon_1\epsilon_2} +\\ \frac{q_\tau^n\hat{y}^{-n}}{2n^3\epsilon_1\beta} + \frac{q_\tau^n\hat{y}^{-n}}{2n^3\epsilon_2\beta} + \frac{q_\tau^n\hat{y}^{-n}}{2n^3\epsilon_1\epsilon_2} + q_\tau^n\hat{y}^{-n}f(n\epsilon_1,n\epsilon_2,n\beta)\frac{1}{n}
\end{eqnarray}
Now it is time to identify the contribution to the error term $E$. The last term 
\begin{equation*}
    \sum_{n\geq 1} q_\tau^n \hat{y}^{-n}f(n\epsilon_1,n\epsilon_2,n\beta)\frac{1}{n}
\end{equation*}
clearly contributes to $E$. The error $E$ also receives contribution from other terms. The numerators $q_\tau^n \hat{y}^{-n}$ in the other four terms is not a function of $m$ only. In \eqref{saddle_region_asymptotic} we want the numerators to be a function of $\hat{m}$ only. So one must perform additional Taylor expansion at $\tau = 0$. This would also contribute to $E$. Let us write it in equation
\begin{equation*}
    E(\epsilon_1,\epsilon_2,\beta,m) = \sum_{n=1}^\infty q_\tau^n \hat{y}^{-n}f(n\epsilon_1, n\epsilon_2,n\beta)\frac{1}{n} + ...
\end{equation*}
where $...$ denotes contribution from the first four terms in \eqref{asymptotic_first_term} which is easy to dealt with. Let us deal with this first. Let us rewrite the sum over the first four terms as polylogarithms.
\begin{eqnarray*}
  \sum_{n=K^{1/4}}^\infty \frac{q_\tau^n\hat{y}^{-n}}{n^4\beta\epsilon_1\epsilon_2} + \frac{q_\tau^n\hat{y}^{-n}}{2n^3\epsilon_1\beta} + \frac{q_\tau^n\hat{y}^{-n}}{2n^3\epsilon_2\beta} + \frac{q_\tau^n\hat{y}^{-n}}{2n^3\epsilon_1\epsilon_2}\\
  = \frac{Li_4(-\beta + \hat{m})}{\beta\epsilon_1\epsilon_2} + \frac{Li_3(-\beta + \hat{m})}{2\beta\epsilon_1}+\frac{Li_3(-\beta + \hat{m})}{2\beta\epsilon_2}+\frac{Li_3(-\beta + \hat{m})}{2\epsilon_2\epsilon_1}\\
  =\frac{Li_4(\hat{m})}{\epsilon_1\epsilon_2\beta} + \frac{Li_3(\hat{m})/2 - Li_4'(\hat{m})}{\epsilon_1\epsilon_2} + \frac{Li_3(\hat{m})}{2\beta}(\frac{1}{\epsilon_1} + \frac{1}{\epsilon_2})+\mathcal{O}_{K/L}(K^{1/4})
\end{eqnarray*}
To go from the first line to the second line we use the definition of the polylogarithms (see the appendix). To go from the second line to the third line we use the fact that the polylogarithm is holomoprhic in a neighborhood of the interval $[i\pi/4, 7 i \pi/4]$.
Therefore one can subtitute in the Taylor expansion
\begin{equation*}
    Li_3(-\beta+\hat{m}) = Li_3(\hat{m}) + \mathcal{O}(\beta), Li_4(-\beta + \hat{m}) = Li_4(\hat{m}) - \beta Li_4'(\hat{m}) + \mathcal{O}(\beta^2)
\end{equation*}
and deduce the $\mathcal{O}_{K/L}(K^{1/4})$ bound which contributes to $E$.

It is equally easy to bound the derivatives of the expression in the first line. If we differentiate with respect to $\hat{m}$, we just need to replace all the polylogarithms in the second and the third line with their derivatives and the error term $\mathcal{O}_{K/L}(K^{1/4})$ remain unchanged. If we differentiate with respect to $\epsilon_1,\epsilon_2,\beta$ increases the bound to $\mathcal{O}_{K/L}(K^{2/4})$. So the contribution from the first four terms on the right hand side in \eqref{asymptotic_first_term} to $E$ satisfies the desired bound \eqref{bound_E}.

Now let us bound
\begin{equation*}
    \sum_{n=1}^\infty q_\tau^n \hat{y}^{-n} f(n\epsilon_1, n\epsilon_2,n\beta)\frac{1}{n}
\end{equation*}
The obvious thing to do is to use the bound on $f$ \eqref{bound_f}. However this bound is only valid when $n|\epsilon_1|<1, n|\epsilon_2|<1, n|\beta| <1$.  We split the sum into two halves so that in the first half the condition 
$n|\epsilon_1|<1, n|\epsilon_2|<1, n|\beta| <1$ is satisfied. In the second half the condition is not satisfied so we will do something else.
\begin{equation*}
    \sum_{n=1}^\infty = \sum_{n = 1}^{1/(200\Re(\beta))} + \sum_{n =1/(200\Re(\beta)) +1}^\infty
\end{equation*}
Here $\Re(\beta)$ refers to the real part in \eqref{real_parts}.
{We} have chosen the number $1/(200\Re(\beta))$ because  $|\Im(\beta)|\leq 100 \Re(\beta)$ so if $n$ is less than this number $n|\beta| < 1$. Since $\Re(\beta)$ is bigger than $\Re(\epsilon_1),\Re(\epsilon_2)$ we also have $n|\epsilon_1|<1, n|\epsilon_2|<1$.

Now we are ready to bound the error term $E$ coming from the sum $n=1... 1/(200\Re(\beta))$. First, using the inequality $|q_\tau^n \hat{y}^{-n}|<1$, we have
\begin{eqnarray*}
    |q_\tau^n\hat{y}^{-n}f(n\epsilon_1,n\epsilon_2,n\beta)| = \mathcal{O}_{K/L}(n^{-1}\epsilon_1^{-1})\\
    \partial_m(q_\tau^n\hat{y}^{-n}f(n\epsilon_1,n\epsilon_2,n\beta)) = n\mathcal{O}_{K/L}(n^{-1}\epsilon_1^{-1})
\end{eqnarray*}
To estimate the derivative in $\beta,\epsilon_1,\epsilon_2$ we use the second equation in \eqref{bound_f}
\begin{equation*}
    \nabla f(\epsilon_1,\epsilon_2,\beta) = \mathcal{O}_{K/L}(\epsilon_1^{-2})\Rightarrow \partial_{\beta,\epsilon_1,\epsilon_2}(f(n\epsilon_1,n\epsilon_2,n\beta)) = n\mathcal{O}_{K/L}(n^{-2}\epsilon_1^{-2})
\end{equation*}
Again, the notation $\partial_{\epsilon_1,\epsilon_2,\beta}A = \mathcal{O}(B)$ means $\partial_\beta A = \mathcal{O}(B), \partial_{\epsilon_1}A = \mathcal{B}, \partial_{\epsilon_2}A = \mathcal{O}(B)$.
Now we include the factor $q_\tau^n\hat{y}^{-n}$:
\begin{equation*}
    \partial_\beta(q_\tau^n\hat{y}^{-n}f(n\epsilon_1,n\epsilon_2,n\beta)) = n\mathcal{O}_{K/L}(n^{-1}\epsilon_1^{-1/4}) + n\mathcal{O}_{K/L}(n^{-2}\epsilon_1^{-2})
\end{equation*}
The first term comes from differentiating the $q_\tau^n$ and the second term comes from differentiating $f$.
Similarly we can bound the derivative with respect to $\epsilon_1,\epsilon_2$
\begin{equation*}
    \partial_{\epsilon_1,\epsilon_2}(q_\tau^n\hat{y}^{-n}f(n\epsilon_1,n\epsilon_2,n\beta)) = n\mathcal{O}_{K/L}(n^{-2}\epsilon_1^{-2})
\end{equation*}
Finally we are ready to write down the bounds for the first half of the sum.
\begin{eqnarray*}
    \sum_{n=1}^{1/(200\Re(\beta))} q_\tau^n\hat{y}^{-n}f(n\epsilon_1,n\epsilon_2,n\beta)\frac{1}{n} = \sum_{n=1}^{1/(200\Re(\beta))}\frac{1}{n}\mathcal{O}_{K/L}(\frac{1}{n\epsilon_1}) = \mathcal{O}_{K/L}(K^{1/4})\\
    \sum_{n=1}^{1/(200\Re(\beta))}\partial_m(q_\tau^n\hat{y}^{-n}f(n\epsilon_1,n\epsilon_2,n\beta)\frac{1}{n}) = \mathcal{O}_{K/L}(K^{1/4}\log K)\\
    \sum_{n=1}^{1/(200\Re(\beta))}\partial_{\beta,\epsilon_1,\epsilon_2}(q_\tau^n\hat{y}^{-n}f(n\epsilon_1,n\epsilon_2,n\beta)\frac{1}{n}) = \mathcal{O}_{K/L}(K^{2/4})
\end{eqnarray*}
where {we} have used the fact that
\begin{equation*}
    \sum_{n=1}^{1/(200\Re(\beta))} \frac{1}{n^2} < \sum_{n=1}^\infty \frac{1}{n^2} < \infty, \sum_{n=1}^{1/(200\Re(\beta))} \frac{1}{n} = \mathcal{O}_{K/L}(\log(K))
\end{equation*}
It is indeed the correct bound on the error term.

Next we bound the error term in the sum from $n=1/(200 \Re(\beta))$ to $\infty$. Instead of estimating the $f$-term directly, let {us} estimate
\begin{equation*}
    \frac{1}{n}\frac{q_\tau^n\hat{y}^{-n}}{(1-q_\tau^n)(1 - q_1^n)(1 - q_2^n)}
\end{equation*}
and
\begin{equation}\label{first_four_terms}
    \frac{q_\tau^n\hat{y}^{-n}}{n^4\beta\epsilon_1\epsilon_2} + \frac{q_\tau^n\hat{y}^{-n}}{2n^3\epsilon_1\beta} + \frac{q_\tau^n\hat{y}^{-n}}{2n^3\epsilon_2\beta} + \frac{q_\tau^n\hat{y}^{-n}}{2n^3\epsilon_1\epsilon_2}
\end{equation}
Since $f$ can be written as the difference of the two we obtain the desired estimates on $f$.
To bound the former, we notice that when $n>1/(200\Re(\beta))$ the real {parts} of $n\epsilon_1, n\epsilon_2,n\beta$ are all bounded below by some positive constants depending on $K/L$. Hence
$(1 - q_\tau^n)^{-1}(1 - q_1^n)^{-1}(1-q_2^n)^{-1} = \mathcal{O}_{K/L}(1)$ is bounded by $\mathcal{O}_{K/L}(1)$ when $n>1/(200\Re(\beta))$.

 so we can write
\begin{eqnarray*}
    \frac{1}{n}\frac{q_\tau^n\hat{y}^{-n}}{(1-q_\tau^n)(1 - q_1^n)(1 - q_2^n)}\\
    = \mathcal{O}_{K/L}(1)K^{-1/4}\sum_{n\geq (1/200\Re(\beta))}|q_\tau \hat{y}^{-1}|^n \\= \mathcal{O}(K/L)K^{-1/4}\frac{|q_\tau \hat{y}^{-1}|^{1/(200\Re(\beta))}}{1 - |q_\tau\hat{y}^{-1}|}
    \\
    = O_{K/L}(1)K^{-1/4}\mathcal{O}_{K/L}(1)K^{1/4} = \mathcal{O}_{K/L}(1)
\end{eqnarray*}
The $K^{-1/4}$ comes from the $1/n$ factor in the first line. To go from the second to the third line we simply sums a geometric series. The go from the third to the last line we use the assumption $|\Re(m)|< 3\Re(\beta)/4$ and hence
\begin{equation*}
    |q_\tau \hat{y}^{-1}|^{1/(200\Re(\beta))} \leq 1, (1 - |q_\tau\hat{y}^{-1}|)^{-1} = \mathcal{O}_{K/L}(K^{1/4})
\end{equation*}
{The estimates for} the derivatives are very similar and we omit the details.

To bound \eqref{first_four_terms}{, we} use the following fact:
\begin{eqnarray*}
   \sum_{n=1/(200\Re(\beta))}^\infty \frac{1}{n^4} = \mathcal{O}_{K/L}(K^{-3/4})\\
   \sum_{n=1/(200\Re(\beta))}^\infty \frac{1}{n^3} = \mathcal{O}_{K/L}(K^{-2/4})
\end{eqnarray*}
Therefore
\begin{equation*}
    \sum_{n=1/(200\Re(\beta))}^\infty \frac{q_\tau^n \hat{y}^{-n}}{n^4\beta\epsilon_1\epsilon_2} \leq \sum_{n=1/(200\Re(\beta))}^\infty \frac{1}{\beta\epsilon_1\epsilon_2 n^4} = \mathcal{O}_{K/L}(K^{-3/4})\mathcal{O}_{K/L}(K^{3/4}) = \mathcal{O}_{K/L}(1)
\end{equation*}
{and} similarly for the other three terms. We can finally write
\begin{equation*}
    \sum_{n=1/(200\Re(\beta))}^\infty \frac{q_\tau^n\hat{y}^{-n}}{n^4\beta\epsilon_1\epsilon_2} + \frac{q_\tau^n\hat{y}^{-n}}{2n^3\epsilon_1\beta} + \frac{q_\tau^n\hat{y}^{-n}}{2n^3\epsilon_2\beta} + \frac{q_\tau^n\hat{y}^{-n}}{2n^3\epsilon_1\epsilon_2} = \mathcal{O}_{K/L}(1)
\end{equation*}
{The estimates for} the derivatives are again similar and we omit {the} details.
To summarize{,} if we write
\begin{align*}
    &\sum_{n\geq 1}\frac{1}{n}\frac{q_\tau^n\hat{y}^{-n}}{(1-q_\tau^n)(1 - q_1^n)(1 - q_2^n)} \\ =&\frac{Li_4(\hat{m})}{\epsilon_1\epsilon_2\beta} + \frac{Li_3(\hat{m})/2 - Li_4'(\hat{m})}{\epsilon_1\epsilon_2} + \frac{Li_3(\hat{m})}{2\beta}(\frac{1}{\epsilon_1} + \frac{1}{\epsilon_2})+E_1(\hat{m},\epsilon_1,\epsilon_2,\beta)
\end{align*}
{then} $E_1$ obeys the same bound as \eqref{bound_E}:
The computation for the other three terms
\begin{eqnarray*}
    \sum_{n\geq 1}\frac{1}{n}\frac{q_\tau^n\hat{y}^{n}q_1^nq_2^n}{(1-q_\tau^n)(1 - q_1^n)(1 - q_2^n)}\\ \sum_{n\geq 1}\frac{1}{n}\frac{q_\tau^nq_1^n}{(1-q_\tau^n)(1 - q_1^n)(1 - q_2^n)}\\
    \sum_{n\geq 1}\frac{1}{n}\frac{q_\tau^nq_2^n}{(1-q_\tau^n)(1 - q_1^n)(1 - q_2^n)}
\end{eqnarray*}
{is} almost identical and we just record the result
\begin{align*}
    &\sum_{n\geq 1}\frac{1}{n}\frac{q_\tau^n\hat{y}^{n}q_1^nq_2^n}{(1-q_\tau^n)(1 - q_1^n)(1 - q_2^n)}
    \\
    =&\frac{Li_4(-\hat{m})}{\beta\epsilon_1\epsilon_2} + \frac{Li_3(-\hat{m})}{-2\beta\epsilon_1}+\frac{Li_3(-\hat{m})}{-2\beta\epsilon_2}+\frac{Li_3(-\hat{m}) - 2Li_4'(-\hat{m})}{2\epsilon_1\epsilon_2} + ...
\end{align*}
And for the other two terms
\begin{eqnarray*}
    \sum_{n\geq 1}\frac{1}{n}\frac{q_\tau^nq_1^n}{(1-q_\tau^n)(1 - q_1^n)(1 - q_2^n)} = \frac{\zeta(4)}{\beta\epsilon_1\epsilon_2} - \frac{\zeta(3)}{2\beta\epsilon_2} + \frac{\zeta(3)}{2\beta\epsilon_1} - \frac{\zeta(3)}{2\epsilon_1\epsilon_2} + ...\\
    \sum_{n\geq 1}\frac{1}{n}\frac{q_\tau^nq_2^n}{(1-q_\tau^n)(1 - q_1^n)(1 - q_2^n)} = \frac{\zeta(4)}{\beta\epsilon_1\epsilon_2} + \frac{\zeta(3)}{2\beta\epsilon_2} - \frac{\zeta(3)}{2\beta\epsilon_1} - \frac{\zeta(3)}{2\epsilon_1\epsilon_2} + ...
\end{eqnarray*}
where $...$ denotes terms which satisfy the same bound as \eqref{bound_E}.
We just add them all up and this proves \eqref{bound_E}. The explicit expressions for the functions $f_i$ are given by:
\begin{equation}\label{f_i_expression}
    f_1(\hat{m}) = f_2(\hat{m}) = \frac{Li_3(\hat{m}) - Li_3(-\hat{m})}{2}, f_3(\hat{m}) = -\frac{Li_3(\hat{m}) + Li_3(-\hat{m})}{2} + \zeta(3)
\end{equation}

\section*{Step 3: Saddle point of the leading asymptotic}
Now we have obtained a good approximation to $\log(Z)$ in the saddle point region, the next task is to find its saddle points. Our previous computation in section 4 shows that there are two saddle points with $|\Im(\epsilon_1)| = \Re(\epsilon_1), |\Im(\epsilon_2)| = \Re(\epsilon_2), |\Im(\beta)| = \Re(\beta)$. In this section we prove a stronger result: the real part of the leading asymptotic
\begin{equation*}
    -\frac{\hat{m}^2(\hat{m}-2\pi i)^2}{24\beta\epsilon_1\epsilon_2} + K\beta + L\epsilon_1 + L\epsilon_2
\end{equation*}
is indeed maximized at the two saddle points. This is what we need since we will {eventually} exponentiate the leading asymptotic and the real part controls the modulus of the exponential. In this section, the domain for $\epsilon_1,\epsilon_2,\beta$ is just the saddle point region:
\begin{equation*}
    |\Im(\epsilon_1)|\leq 100 \Re(\epsilon_1), |\Im(\epsilon_2)|\leq 100 \Re(\epsilon_2),|\Im(\beta)|\leq 100 \Re(\beta)
\end{equation*}
However we allow the domain for $\hat{m}$ to be slightly bigger $\hat{m}\in [0, 2\pi i]$ to make our calculation simpler. If we can maximize the real part over a larger domain and the maximum lies in the smaller domain, then we would know the maximum in the smaller domain.

Before proceeding with our computation{,} let us introduce some variables $\sigma_i, i=1,2,3,4$ which will be order 1 in the large $K$ limit:
\begin{equation*}
    \epsilon_1 = \Re(\epsilon_1)\sigma_1, \epsilon_2 = \Re(\epsilon_2)\sigma_2, \beta = \Re(\beta)\sigma_3, \hat{m} = \sigma_4
\end{equation*}
These order 1 variables will {make} this computation easier. There is a Jacobian factor coming from change of variable in the integral. The Jacobian is a monomial in $K,L$ and can be absorbed into $P$ in \eqref{asymptotic}. So from now on we ignore this Jacobian. Using these new variables we can write
\begin{align}\label{leading_asymptotic_new_variables}
    &-\frac{\hat{m}^2(\hat{m}-2\pi i)^2}{24\beta\epsilon_1\epsilon_2} + K\beta + L\epsilon_1 + L\epsilon_2\nonumber\\
    =& C(K,L) \left(-\frac{\hat{m}^2(\hat{m}-2 \pi i)^2}{24\sigma_1\sigma_2\sigma_3} + \frac{\pi^4}{96}(\sigma_1 + \sigma_2 + \sigma_3)\right)\nonumber\\
    \coloneqq& C(K,L)S(\sigma_1,\sigma_2,\sigma_3,\sigma_4 = \hat{m})
\end{align}
where
\begin{equation*}
    C(K,L) = (\Re(\beta)\Re(\epsilon_1)\Re(\epsilon_2))^{-1} = \frac{6^{3/4}K^{1/4}8\sqrt{L}}{\pi^3}
\end{equation*}

The domain for $\sigma_1, \sigma_2, \sigma_3, \sigma_4$ is $[1 - 100i, 1 + 100i]^3\times [0, 2\pi i]$. This is a compact set so $\Re (S)$ attains its maximum (possibly at more than one point) either on the boundary or in the interior.  We first show that $\Re (S)$ cannot attain its maximum on the boundary by comparing its boundary value with $\Re (S)(\sigma^\pm) = \pi^4/24$.
The boundary of the domain $[1 - 100i, 1 + 100i]^3\times [0, 2\pi i]$ {is the union of four pieces, each of which is the product of an endpoint of an interval and the other three intervals}. If $\hat{m} = 0, 2\pi i$ then $\Re(S) = \pi^4/36 < \pi^4/24$. When $\Im(\sigma_1) = 100 \Re(\sigma_1)$, $\Re(S)\leq \pi^4/36 + \pi^4/(100 \times 24) < \pi^4/24$ and similarly for $\sigma_2,\sigma_3$. So $\Re S$ cannot attain its maximum on the boundary.

To compute the maximum in the interior, let us differentiate with respect to $\sigma_1, \sigma_2,\sigma_3, \sigma_4$ along purely imaginary directions and set the real part of the derivatives to be zero. Differentiating with respect to the imaginary direction is the same as $i$ times the usual holomorphic derivative. So we set 
\begin{equation*}
    {\Re(i\partial_{\sigma_1}S) = \Re(i\partial_{\sigma_2}S) = \Re(i\partial_{\sigma_3}S) = \Re(i\partial_{\hat{m}}S) = 0}
\end{equation*}
We will postpone the discussion of $\hat{m}$-derivative. Let us look at the derivative with respect to $\sigma_1,\sigma_2,\sigma_3$ first:
which implies
\begin{equation*}
    \Im(\frac{\hat{m}^2(\hat{m} - 2\pi i)^2}{\sigma_1^2\sigma_2\sigma_3}) = \Im(\frac{\hat{m}^2(\hat{m} - 2\pi i)^2}{\sigma_1\sigma_2^2\sigma_3}) = \Im(\frac{\hat{m}^2(\hat{m} - 2\pi i)^2}{\sigma_1\sigma_2\sigma_3^2})=0
\end{equation*}
Hence $\sigma_1^2\sigma_2\sigma_3, \sigma_1\sigma_2^2\sigma_3,\sigma_1\sigma_2\sigma_3^2$ are all real. They are nonzero since they are products of nonzero numbers. So the ratios $\sigma_1/\sigma_2, \sigma_2/\sigma_3$ {are} both real. Since $\sigma_1, \sigma_2,\sigma_3$ have the same real part, {we have} $\sigma_1=\sigma_2=\sigma_3$. Since $\sigma_1^2\sigma_2\sigma_3$ is real, there are only three possible solutions: $\sigma_1 = \sigma_2 = \sigma_3 = 1+ i,1-i, 1$. The solution $\sigma_1 = \sigma_2 = \sigma_3= 1$ does not maximize the real part because $\Re (S) < \pi^2/32$ at that point. It remains to consider $\sigma_1 = \sigma_2 = \sigma_3 = 1\pm i$. Now we maximize with respect to $\hat{m}$. Since $\hat{m}^2(\hat{m} - 2\pi i)^2$ is maximized at $\hat{m} = i\pi$ so $\Re(S)$ is maximized at  $\sigma_1 = \sigma_2 = \sigma_3 = 1\pm i, \hat{m} = i\pi$.
The two saddle points are complex conjugate to each other. (For $\hat{m}$ one needs to add $2\pi i$ after complex conjugation).

The only remaining task in this section is to compute the Hessian $\partial_{\sigma_i}\partial_{\sigma_j}S(\sigma^\pm)$and shows that it is negative along purely imaginary directions. We just record the result: (the columns are from left to right: $\sigma_1,\sigma_2,\sigma_3, \sigma_4 = \hat{m}$)
\begin{equation}\label{hessian}
    H[\sigma_i^+] =  -\frac{\pi^4}{24(1+i)^5}\begin{bmatrix}
    2&1&1&0\\
    1&2&1&0\\
    1&1&2&0\\
    0&0&0&8i/\pi^2\\
    \end{bmatrix}, H[\sigma_i^-] = \overline{H[\sigma_i^+]}
\end{equation}
And we see that locally the second derivative of the real part is negative definite along the four contours:
\begin{equation*}
    \Re(iv^T H(\sigma_i^+) iv) < 0, \forall v\in \R^4\neq 0
\end{equation*}
where the purely imaginary column vector $iv$ represents fluctuation of $\sigma_i$ along the four contours $[1-100i, 1+ 100i]^3\times [0,2\pi i]$. The negative definiteness follows because the metrix $[2,1,1;1,2,1;1,1,2]$ is positive definite (its eigenvalues are $4,1,1$) and the real part of $-(1+i)^{-5}$ is positive. For the $\hat{m}$-component, the real part of $-(\pi^4/(24(1+i)^5))(8i/\pi^2)$ is positive.

\section*{Step 4: Saddle point of the leading and the first subleading asymptotic}
At this point one might be tempted to perform quadratic fluctuation of $\log(Z)$ around the two approximate saddle points $\sigma_i^\pm$. It does not work at this stage because $\log(Z) + K\beta + L\epsilon_1 + L\epsilon_2$ has a nonzero first derivative at $\sigma_i^\pm$ and the nonzero first derivative leads to difficulties if one tries to compute Guassian integrals around the two saddle points. To see why,
let us write the integrand as
\begin{equation}\label{F_definition}
    \exp\left(C(K,L)F_{K,L}(\sigma_i))\right), F_{K,L}(\sigma_1,\sigma_2,\sigma_3,\sigma_4) = \frac{\log(Z) + K\beta + L\epsilon_1 + L\epsilon_2}{C(K,L)}
\end{equation}
The function $F$ depends on $K,L$ but we will soon suppress this dependence to make the notation less cumbersome. The notation $\sigma_i$ means $\sigma_1,\sigma_2,\sigma_3,\sigma_4$. In the limit $K\to \infty$
\begin{equation*}
F_{K,L}(\sigma_1,\sigma_2,\sigma_3,\sigma_4)\to S(\sigma_i) = -\frac{\hat{m}^2(\hat{m} - 2\pi i)^2}{24\sigma_1\sigma_2\sigma_3} + \frac{\pi^2}{96}(\sigma_1 + \sigma_2 + \sigma_3)
\end{equation*}

From now we will not write the subscripts in $F$ but one must keep in mind that $F$ depends on $K,L$. We perform a Taylor expansion around the saddle point $\sigma^+$ \begin{equation*}
    F(\sigma_i) = F(\sigma^+) + \partial_iF \delta\sigma_i + \frac{1}{2}\partial_i\partial_jF \delta\sigma_i\delta\sigma_j+..., \sigma_i = \sigma^+_i +\delta\sigma_i
\end{equation*}
The function $F$ can be expanded in a Taylor series in $K^{-1/4}$. The zeroth order term is the function $S$ {defined in }\eqref{leading_asymptotic_new_variables} which  has zero derivative at $\sigma^\pm$. However the order $K^{-1/4}$ term has non-vanishing derivative at $\sigma^\pm$ so $\partial_iF$ is of order $K^{-1/4}$. When we compute the Guassian fluctuation around the saddle point $\sigma^+$ we need to define new variables to absorb the $C(K,L)$ into the second derivative. In other words{,} we define
\begin{equation*}
    \chi_i = \sqrt{C(K,L)}\delta \sigma_i\sim K^{3/8}\delta\sigma_i
\end{equation*}
The integrand is written schematically as
\begin{eqnarray*}
    \exp\left(K^{3/4}(F(\sigma^+) + K^{-1/4}\delta \sigma - \delta\sigma\delta\sigma-...)\right) \\= \exp\left(K^{3/4}F(\sigma^+) + K^{1/8}\chi - \chi\chi+...\right)
\end{eqnarray*}
where I have omitted the various constant factors wchih depends only on $L/K$. When we perform the Gaussian integral
\begin{equation*}
    \int d\chi_1d\chi_2d\chi_3d\chi_4\exp(K^{1/8}\chi - \chi\chi+...)
\end{equation*}
The divergent linear factor $K^{1/8}\chi$ means one cannot naively apply Lebesgue dominated convergence to get what we want.

Therefore we will slightly shift the saddle points $\sigma^\pm$ to a new set of saddle points $\tilde{\sigma}^\pm$ and the derivatives of $\log(Z) + K\beta +L \epsilon_1 + L\epsilon_2$ at $\tilde{\sigma}^\pm$ do not grow as fast as the derivatives at $\sigma^\pm$. To compute the lcocation of $\tilde{\sigma}^\pm$ we need to use subleading terms in $\log(Z)$: In other words $\tilde{\sigma}^\pm$ are the stationary points of 
\begin{equation}\label{subleading}
    \tilde{S} = -\frac{\hat{m}^2(\hat{m}-2\pi i)^2}{24\beta\epsilon_1\epsilon_2} + K\beta + L\epsilon_1 + L\epsilon_2 + \frac{f_3(\hat{m})}{\epsilon_1\epsilon_2} + \frac{f_2(\hat{m})}{\beta_1\epsilon_2}+\frac{f_1(\hat{m})}{\beta\epsilon_1}
\end{equation}
where I have included the first subleading corrections. I have also temporarily switched back to the original set of variables $\epsilon_1,\epsilon_2,\beta,\hat{m}$ to make the computation easier. The change of variable formula \eqref{change_of_variables} remains the same. After writing everything using the new variables $\sigma_i$ the three subleading terms decay like $K^{-1/4}$ relative to the order one term. And hence the implicit function theorem implies that $\tilde{\sigma}^\pm$ exist when $K$ is large and converges to $\sigma^\pm$ as $K\to \infty$. Moreover when $K$ is sufficiently large the difference $\tilde{\sigma}^\pm - \sigma^\pm$ can be written as a power series in $K^{-1/4}$. This fact will be important when we prove the polynomiality of` subleading terms in \eqref{asymptotic}. The news saddle points $\tilde{\sigma}^\pm$ can be written using $\epsilon_1,\epsilon_2,\beta,\hat{m}$ or $\sigma_1,\sigma_2,\sigma_3,\sigma_4$.

Proving the existence of $\tilde{\sigma}^\pm$ is not enough. We still need to deform the contours to pass through $\tilde{\sigma}^\pm$. Therefore, we must ensure that $\tilde{\sigma}^\pm$ lie in the domain of definition $|\Re(m)|<3\Re(\beta)/4$. So let us compute {$\tilde{\sigma}^\pm_4$}, the new saddle point {location} for $\hat{m}$.
We write $\hat{m} = i\pi + \delta \hat{m}$ and differentiate with respect to $\delta \hat{m}$. We anticipate that the new saddle point $\tilde{\sigma}^\pm$ is attained at $\delta \hat{m}$ small but nonzero so we perform Taylor expansion of \eqref{subleading} at $\hat{m} = i\pi$ and keep only the leading order terms in $\delta \hat{m}$. For example we write $\hat{m}^2(\hat{m} - 2\pi i)^2 = \pi^4 + 2\pi^2\delta \hat{m}^2 + \mathcal{O}(\delta \hat{m}^4)$. When we differentiate with respect to $\delta \hat{m}$ in this term we would get a factor of $\delta \hat{m}/(\beta\epsilon_1\epsilon_2)\sim K^{2/4}$. 

We also need to linearise the three subleading terms containing $f_1,f_2,f_3$ at $\hat{m} = i\pi$ and keep terms which {grow} like $K^{2/4}$ after differentiating with respect to $\delta \hat{m}$. For $f_3/(\epsilon_2\epsilon_1)$ the whole term can be discarded because $f_3'(i\pi) = 0$ and hence the term can be approximated (up to a term independent of $\hat{m}$) as  $f_3''(i\pi)\delta \hat{m}^2/(2\epsilon_2\epsilon_1)$. if we differentiate with respect to $\delta \hat{m}$ the derivative would grow at most {like} $K^{1/4}$ and so can be safely discarded. The other two subleading terms cannot be discarded and we arrive at the following equation for $\delta \hat{m}$ at $\sigma^\pm$:
\begin{equation*}
    \frac{-4\pi^2\delta \hat{m}}{24\beta\epsilon_1\epsilon_2} + \frac{f_1'(i\pi)}{\beta\epsilon_1} + \frac{f_2'(i\pi)}{\beta\epsilon_2} = 0
\end{equation*}
{which implies}
\begin{equation*}
    \tilde{\sigma}_4^\pm = i\pi - \frac{\epsilon_1 + \epsilon_2}{2} + \mathcal{O}_{K/L}(K^{-2/4})
\end{equation*}
{Here} $\epsilon_1,\epsilon_2$ refer to the leading saddle point values of { $\epsilon_1,\epsilon_2$}:
\begin{equation*}
    \epsilon_1 = \Re(\epsilon_1)\pm i\Re(\epsilon_1), \epsilon_2 = \Re(\epsilon_2)\pm i\Re(\epsilon_2) = \epsilon_1
\end{equation*}
and $\Re(\epsilon_1),\Re(\epsilon_2)$ are given by \eqref{real_parts} and the choice of $\pm$ agrees with the choice of $\pm$ in $\tilde{\sigma}^\pm$.
Hence if we set $\Re(\epsilon_1) < \Re(\beta)/2$ this saddle point value of $\hat{m}$ lies in the domain of definition $|\Re(m)| < 3\Re(\beta)/4$. We also note that $\tilde{\sigma}^\pm$ are complex conjugate to each other since $\tilde{S}$ in\eqref{subleading} satisfies
\begin{equation*}
    \tilde{S}[\bar{\beta},\bar{\epsilon}_1,\bar{\epsilon}_2,\bar{\hat{m}}] = \overline{\tilde{S}[\beta,\epsilon_1,\epsilon_2,\hat{m}]}
\end{equation*}
And the complex conjugation of a stationary point is also a stationary point. We have achieved our goal for this section: the first derivative of $\log(Z) + L\epsilon_1 + L\epsilon_2 + K\beta$ at $\tilde{\sigma}^\pm$ receives contribution from $E$ only and satisfies the bound \eqref{bound_E}. Hence and
the derivatives of $F$ are bounded by
\begin{equation*}
    \partial_{\sigma_1, \sigma_2, \sigma_3 ,\hat{m} = \sigma_4}F(\tilde{\sigma}^\pm) = \mathcal{O}_{K/L}(K^{-2/4}\log(E))
\end{equation*}
So we have improved the bound on $\nabla_{\sigma_i} F$ from $K^{-1/4}$ to $K^{-2/4}\log(K)$ by shifting the saddle points to $\tilde{\sigma}^\pm$.

\section*{Step 5: Computation around the saddle points}
Now we have two new {saddle points} $\tilde{\sigma}^\pm$ so one can deform the four contours {slightly to} pass through $\tilde{\sigma}^\pm$. In other words, we deform the contour of $\hat{m}$ to pass through $\tilde{\sigma}_4^\pm$, we deform the contour of $\epsilon_1$ to pass through $\tilde{\sigma}_1^\pm$ and similarly for the other two variables. The {contours} for $\sigma_i$ are also deformed according to \eqref{change_of_variables}. From now on we will exclusively use the variables $\sigma_i$. To be more precise: choose a smooth function $\phi$ defined on $[7i\pi/4, i\pi/4]$ so that $\phi = 0$ in a neighborhood of $i\pi/4$ and a neighborhood of $7i\pi/4$ and $\phi = 1$ in a neighborhood of $i\pi$. {In addition, }$\phi$ is monotonic on the two intervals on which it smoothly connects $0$ and $1$. The contours for $\hat{m}$ is parametrized by $t\mapsto t + \phi(t)\Re(\sigma_4^+)$ for $t\in[i\pi/4,7i\pi/4]$.

The next step in the saddle point method is to zoom in small neighborhoods of the saddle points and perform Gaussian integrals in the two neighborhoods. {First,} let us specify the neighborhoods:
\begin{equation*}
    \tilde{B}^\pm := \{\sum_{i=1}^4|\sigma_i - \tilde{\sigma}_i^\pm|^2 < K^{-1/50}\}
\end{equation*}
and we require that $\sigma_i$ lie on the deformed contour.
So we have two balls $\tilde{B}^\pm$ centred at the two saddle points $\tilde{\sigma}^\pm$ and the radii of the two balls tend to zero very slowly like $K^{-1/100}$. Here is why: we will soon approximate the integrand $Z$ in the two balls by its Taylor expansion to second order. If the radii do not tend to zero, the approximation does not get better as $K\to \infty$. On the other hand, if the radii go to zero too rapidly we would not be able to bound the error away from the two balls. The power $-1/50$ can be replaced by any other number provided it is sufficiently small.

Before {computing the Gaussian fluctuations inside the two balls,} let us first bound the {errors} away from the two balls. Let us write down the integral again using {the} variables $\sigma_i$:
\begin{equation*}
    \int d\sigma_i \exp\left(C(K,L)(S(\sigma_i) + S_{-1/4}(\sigma_i) + S_{-2/4}(\sigma_i))\right)
\end{equation*}
where
\begin{align*}
    &F = S + S_{-1/4} + S_{-2/4}\\
    &S(\sigma_i) = -\frac{\hat{m}^2(\hat{m}-2\pi i)^2}{24\sigma_1\sigma_2\sigma_3} + \frac{\pi^2}{96}(\sigma_1 + \sigma_2 + \sigma_3)\\
    &S_{-1/4}(\sigma_i) = \Re(\epsilon_2)\frac{f_2(\sigma_4)}{\sigma_3\sigma_1} + \Re(\epsilon_1)\frac{f_1(\sigma_4)}{\sigma_3\sigma_2} + \Re(\beta)\frac{f3(\sigma_4)}{\sigma_1\sigma_2}\\
    &S_{-2/4} = \frac{E(\Re(\epsilon_1)\sigma_1, \Re(\epsilon_2)\sigma_2,\Re(\beta)\beta, \sigma_4)}{C(K,L)}
\end{align*}
The notation $S(\sigma_i)$ means we regard $S$ as a function of $\sigma_1,\sigma_2,\sigma_3,\sigma_4$. The measure $d\sigma_i$ means $d\sigma_1d\sigma_2d\sigma_3d\sigma_4$.
Again we have omitted the factor of $2\pi i$ outside the integral and the Jacobian factor coming from change of variable to $\sigma_i$. Both of them can absorbed into the function $P$ in \eqref{asymptotic}. The subscripts indicate the decay rate as $K\to\infty $. $S(\sigma_i)$ does not decay. $S_{-1/4}$ decays like $K^{-1/4}$ and $S_{-2/4}= \mathcal{O}_{K/L}(K^{-2/4})$.

{The} predicted asymptotic is $\exp(\#K^{3/4})$ ($\#$ denotes some positive number) so when we bound the growth away from the two balls we can ignore {any term that grows} like $\exp(\#K^{2/4})$ or slower. In particular{,}  we can ignore $S_{-1/4}$ and $S_{-2/4}$ in this estimate. So it remains to bound $S$ on the complement of the two balls $\tilde{B}^\pm$. In this region {we} will show that $\Re (S)$ is bounded by its {value} on $\partial \tilde{B}^\pm$ and then use the negative definite of the Hessian to produce a bound on $S$. We have shown in one of the previous sections that on the original {contours} $\Re (S)$ has vanishing derivatives at only two points. This remains true when the {contours} are slightly perturbed. Since everything is analytic in $K^{-1/4}$, the implicit function theorem implies that the two stationary points of $\Re (S)$ would converge to $\sigma^\pm$ like $K^{-1/4}$. In particular when $K$ is large the two new stationary points would lie inside $\tilde{B}^\pm$. As a result there is no stationary point for $\Re (S)$ in the complement of $B^\pm$. Hence in the complement of $\tilde{B}^\pm$, $\Re (S)$ attains its maximum either {on} $\partial \tilde{B}^\pm$ or when one of the variables {reaches} the boundary of its contour. Our analysis in {Step 3} (after only a minor change because now the contour for $\hat{m}$ is $[i\pi, 7i\pi/4]$) shows that the second scenario does not occur when $K$ is sufficiently large. Hence $\Re (S)$ attains its maximum on $\partial \tilde{B}^\pm$. To estimate $\Re (S)$ on $\partial \tilde{B}^\pm$ we use the negative definiteness of the real part of the Hessian. Actually, it is easier to estimate $\Re (S)$ on $\partial B^\pm$ where
\begin{equation*}
   B^\pm:= \{\sum_{i=1}^4|\sigma_i - \tilde{\sigma}_i^\pm|^2 < K^{-1/50}\}, \Re(\sigma_4) = 0, \Re(\sigma_1) = \Re(\sigma_2) = \Re(\sigma_3) = 1
\end{equation*}
$\tilde B^\pm \to B^\pm$ as $K\to\infty$. The difference between $\Re (S)$ on $\partial \tilde{B}^\pm$ and $\partial B^\pm$ is of order $K^{-1/4}$ so it sufficies to estimate $\Re (S)$ on $\partial B^\pm$.
\begin{equation*}
    \Re (S)(\partial B^\pm)\leq \Re (S)(\sigma^\pm) - \frac{1}{2}\lambda(K^{-1/100})^2
\end{equation*}
where $\lambda >0$ is a bit less than the smallest eigenvalue of minus the real part of the Hessian of $S$.
The predicted leading order growth is {$\exp(C(K,L)\Re(S)(\sigma^\pm))$ }and therefore 
\begin{equation*}
    \frac{\exp(C(K,L)\Re (S)(\partial B^\pm))}{ \exp(C(K,L)\Re (S)(\sigma^\pm))} \leq \exp(C(K,L)(-1/2)\lambda K^{-1/50})
\end{equation*}
decays exponentially like $\exp(-\#K^{3-1/50})$ and is clearly $o(1)$.

Finally we can restrict our contour integral to just the two balls $\tilde{B}^\pm$. We will study the case of $\tilde{B}^+$ and the other case $\tilde{B}^-$ is completely analogous. We perform a change of variables
\begin{equation*}
    \sigma_i = \tilde{\sigma}_i + \frac{\chi_i} {\sqrt{C(K,L)}}, \sum_{i=1}^4 \chi_i^2 < C(K,L) K^{-1/50}
\end{equation*}
So {the} radius square of the domain of integration of $\chi_i$ grows like $K^{3-1/50}$ and tends to infinity as $K\to \infty$.

It remains to approximate $F$ by its Taylor expansion to second order in the two balls. Again we do the case of $\tilde{\sigma}^+$ first.
\begin{eqnarray*}
    F(\sigma_i) = F(\tilde{\sigma}^+) + \partial_iF(\tilde{\sigma}^+) \frac{\chi_i}{\sqrt{C(K,L)}} + \frac{1}{2}\partial_i\partial_jF(\tilde{\sigma}^+) \frac{\chi_i\chi_j}{C(K,L)} + \frac{\sup_{\tilde{B}^+}(\nabla^3 F)\mathcal{O}_{K/L}(\chi^3)}{C(K,L)^{3/2}}
\end{eqnarray*}
After we multiply both sides by $C(K,L)${,} the first derivative term goes to zero pointwise in $\chi$ since $\partial_i F = \mathcal{O}_{K/L}(K^{-2/4}\log K)$ and the third derivative term goes to zero provided we can bound $\nabla^3F$  over the whole ball $\tilde{B}^+$. We also need to show that the second derivative term converges pointwise. These two estimates use the same technique developed in the previous section and the reader can refer to the appendix for more details.

So now we can rewrite the integral in the ball $B^+$
\begin{equation*}
    I(K,L)\coloneqq\int d\chi \exp(C(K,L)F(\sigma_i)) = \int d\chi \exp(C(K,L)F(\tilde{\sigma}^+) + \frac{1}{2}\partial_i\partial_kF(\tilde{\sigma}^+)\chi_i\chi_j + ...)
\end{equation*}
where the measure $d\chi = d\chi_1d\chi_2d\chi_3d\chi_4$ where $...$ denotes terms which converge to zero pointwise in $\chi$.

We apply Lebesgue dominated convergence to
\begin{equation}\label{gaussian_integral}
    \int d\chi \exp(\frac{1}{2}\partial_i\partial_jF(\tilde{\sigma}^+)\chi_i\chi_j + ...)
\end{equation}
to conclude
\begin{equation*}
    I(K,L) = \exp(C(K,L)F(\tilde{\sigma}^+))(C + o(1))
\end{equation*}
where the constant $C$ is the Gaussian integral
\begin{equation}
    \int d\chi \exp(\frac{1}{2}H(\sigma^+)\chi_i\chi_j)
\end{equation}
where $H$ is given by \eqref{hessian}.
The negative definiteness of the Hessian guarantees the existence of the integral. The dominating function we used is 
\begin{equation}
    |\exp(\frac{1}{4} H(\sigma^+)\chi_i\chi_j)|
\end{equation}
The object $I(K,L)$ equals $\mathcal{C}(2L,0,0,K)$ times a suitable power of $2\pi i$ and a Jacobian which is a monomial in $K,L$. Hence to prove \eqref{asymptotic}{,} it suffices to prove that the saddle point value $F(\tilde{\sigma}^\pm)$ are polynomials in $K,\log(K)$.

\section*{Step 6: Asymptotic of the saddle point value}
The only remaining task in this paper is to compute the asymptotic value of the integrand at the saddle points $\exp(F(\tilde{\sigma}^+))$ as $K\to \infty$. We will only analyze the saddle point $\tilde{\sigma}^+$ and $\tilde{\sigma}^-$ is similar. The following method is originally due to Newman \cite{Newman}. In this section we switch to the original variables $\epsilon_1,\epsilon_2,\beta,\hat{m}$. Let
\begin{equation*}
    \delta = K^{-1/4}, \epsilon_1 = R_1\delta, \epsilon_2 = R_2\delta, \beta = R_3\delta, \hat{m} = i\pi + R_4\delta
\end{equation*}
At $\tilde{\sigma}^+$, $R_1,R_2,R_3,R_4$ are all bounded below and above by positive constants depending on $K/L$. It turns out that terms which include $\hat{y}$ and terms which do not include $\hat{y}$ have different asymptotic so we need to study {them} separately:
\begin{equation*}
    \log(Z) = \underbrace{\sum_{n\geq 1}\frac{1}{n}\frac{q_\tau^n(q_1^n + q_2^n)}{(1-q_\tau^n)(1-q_1^n)(1-q_2^n)}}_{F_1} + \underbrace{\sum_{n\geq 1}\frac{1}{n} \frac{q_\tau^n\hat{y}^{-n} + q_\tau^n \hat{y}^nq_1^nq_2^n}{(1- q_\tau^n)(1 - q_1^n)(1 - q_2^n)}}_{F_2}
\end{equation*}
We will study $F_1$ first. We write $F_1$ as a function of $\delta, R_1,R_2,R_3,R_4$
\begin{equation*}
    F_1= \delta \sum_{n\geq 1}\frac{1}{n\delta}\frac{\exp(-(R_3 + R_1)n\delta) +\exp(-(R_3 + R_2)n\delta)}{(1 - \exp(-nR_3\delta))(1 - \exp(-nR_1\delta))(1 - \exp(-nR_2\delta))}
\end{equation*}
This is the Riemann sum of the following function:
\begin{equation*}
    G_1(x) = \frac{1}{x}\frac{\exp(-(R_3+ R_1)x)+\exp((R_3 + R_2)x)}{(1 - \exp(-R_3x))(1 - \exp(-R_1x))(1 - \exp(-R_2x))}
\end{equation*}
on the interval $[0,\infty)$. We will approximate the Riemann sum $
F_1$ by the Riemann sum of another function $\tilde{G}_1$:
\begin{equation*}
\tilde{G}_1(x) = \frac{P_1(R_i)}{x^4} + \frac{P_2(R_i)}{x^3} + \frac{P_3(R_i)}{x^2} + \frac{\exp(-x)P_4(R_i)}{x}    
\end{equation*}
The notation $P_1(R_i)$ means $P_1(R_1, R_2,R_3,R_4)$.
Its Riemann sum is
\begin{equation}\label{approximate_riemann_sum_1}
    \delta \sum_{n\geq 1}\frac{P_1(R_i)}{n^4\delta^4} + \frac{P_2(R_i)}{n^3\delta^3} + \frac{P_3(R_i)}{n^2\delta^2} + \frac{\exp(-n\delta)P_4(R_i)}{n\delta}
\end{equation}
The coefficients $P_1,P_2,P_3,P_4$ are given by the Taylor expansion of $G_1$ at $x = 0$:
\begin{equation*}
    G_1(x) =  \frac{P_1(R_i)}{x^4} + \frac{P_2(R_i)}{x^3} + \frac{P_3(R_i)}{x^2} + \frac{P_4(R_i)}{x}+ \mathcal{O}_{K/L}(1), 0<x<1
\end{equation*}
This choice of $P_i$ ensures {the regularity of} $G_1 - \tilde{G_1}$ at $x = 0$. Since both $G_1$ and $\tilde{G_1}$ decay at least like $x^{-2}$ when $x\to \infty$. Their difference $G_1 - \tilde{G_1}$ is integrable on $[0,\infty)$ and the difference of their Riemann sums tends to zero. The approximate Riemann sum \eqref{approximate_riemann_sum_1} can be rewritten as a linear sum of $\delta^{-3},\delta^{-2},\delta,\log(\delta)$ and $o(1)$. Here is why: the $R_i$ all tend to their limiting values when $\delta\to 0$ and can be expanded as a Taylor series in $\delta$. The term
\begin{equation*}
    \sum_{n\geq 1}\frac{\exp(-n\delta)}{n} = -\log(1 - \exp(-\delta)) = -\log(\delta) + \mathcal{O}(1)
\end{equation*}
This proves that $F_1$ can be expressed as a linear sum of $\delta^{-3},\delta^{-2},\delta,\log(\delta), 1, o(1)$.

The similar method works for $F_2$. However there is one important difference. When we write $F_2 $ in terms of $\delta,R_1,R_2,R_3,R_4$ we get an extra factor of $(-1)^n$:
\begin{equation*}
F_2 = \delta \sum_{n\geq 1}\frac{1}{n\delta}(-1)^n\frac{\exp(-(R_3 - R_4)n\delta)+\exp(-(R_3 + R_4 + R_1 + R_2)n\delta)}{(1 - \exp(-nR_3\delta))(1 - \exp(-nR_1\delta))(1 - \exp(-nR_2\delta))}
\end{equation*}
Due to the $(-1)^n$ this is not a Riemann sum but a difference between two Riemann sums corresponding to $n$ even/odd. We approximate $F_2$ by
\begin{equation}\label{approximate_riemann_sum_2}
    \delta \sum_{n\geq 1}(-1)^n\left(\frac{Q_1(R_i)}{n^4\delta^4} + \frac{Q_2(R_i)}{n^3\delta^3} + \frac{Q_3(R_i)}{n^2\delta^2} + \frac{\exp(-n\delta)Q_4(R_i)}{n\delta}\right)
\end{equation}
where $Q_i$ are determined by the Taylor expansion of
\begin{equation*}
    G_2(x) = \frac{1}{x}\frac{\exp(-(R_3- R_4)x)+\exp(-(R_1 + R_2 + R_3 + R_4)x)}{(1 - \exp(-R_3x))(1 - \exp(-R_1x))(1 - \exp(-R_2x))}
\end{equation*}
Its Taylor expansion at $x = 0$ is
\begin{equation}
    G_2(x) = \frac{Q_1(R_i)}{x^4}+\frac{Q_2(R_i)}{x^3}+\frac{Q_3(R_i)}{x^2} + \frac{Q_4(R_i)}{x}+ \mathcal{O}_{K/L}(1)
\end{equation}
We have $F_2$ - \eqref{approximate_riemann_sum_2} tends to zero. Due to the $(-1)^n$ \eqref{approximate_riemann_sum_2} can be written as a sum of $\delta^{-3},\delta^{-2},\delta^{-1},1, o(1)$ and there is no $\log(\delta)$ term because
\begin{equation}
    \sum_{n\geq 1}(-1)^n\frac{\exp(-n\delta)}{n\delta} = - \log(1 + \exp(-\delta))
\end{equation}
 is smooth at $\delta = 0$. In summary the saddle point value $\exp(C(K,L)F(\tilde{\sigma}^+)$ equals:
 \begin{equation}
     \exp(2\sqrt{2}\pi 24^{-1/4} \sqrt{L}K^{1/4}(1 + i) + A_2 K^{-2/4} + A_3 K^{-1/4} + A_4 \log(K) + A_5 + o(1))
 \end{equation}
 for some complex numbers $A_2,A_3,A_4,A_5$. {The two balls contribute complex conjugate saddle point values}. Hence we just need to take twice the real part. Absorbing the $(2\pi i)^{-4}$ into $A_5$ and the Jacobian into $A_4$ we finally obtain the desired conclusion (up to an overall factor of $2$ which can be {absorbed} into the exponential)
 \begin{equation}
     C(2K, 0, 0, L) = \exp(2\sqrt{2}\pi 24^{-1/4} \sqrt{L}K^{1/4}+...)(\cos(2\sqrt{2}\pi 24^{-1/4} \sqrt{L}K^{1/4}+...) + o(1))
 \end{equation}
where the two $...$ {denote} the real and imaginary parts of $A_2 K^{-2/4} + A_3 K^{-1/4} + A_4 \log(K) + A_5$ respectively.

\section*{Computational results}
In this section we present and explain the numerical data for $\mathcal{C}(L, 0, 0, K)$
\begin{center}
\begin{tabular}{|c|c|}
\hline
$(2L, K)$ & $\mathcal{C}(2L,0,0,K)$\\
\hline
(24, 25) & -4548136426 \\
(26, 26) & 7935209206\\
(28, 27) & 88800402896\\
(30, 28) & -4887654890 \\
(32, 29) & -1425581403152\\
(40, 45) & 7337290205677620\\
(42, 46) & 23800263998384620\\
(44, 47) & -51625798313702826 \\
(46, 48) & -429211479407800616 \\
(48, 49) & -288354194415296772 \\
(50, 50) & 4773158006473089778 \\
(52, 51) & 14870285533157146362\\
(54, 52) & -21630735101481854366 \\
\hline
\end{tabular}
\end{center}
IF we only take the leading order term in \eqref{asymptotic}, the convergence is slow. So we will take into account subleading corrections. The analysis in the previous section shows that subleading correction of $\log|\mathcal{C}|$ can take {$K^{2/4}, K^{1/4}, \log(K)$ or $\mathcal{O}_{K/L}(1)$}. For small $K$ it turns out {that} $\log(K)$ contributes more than the other terms. Part of the contribution from $\log(K)$ comes from the quadratic determinant which is easy to compute. The determinant of the second derivative of \eqref{leading_exponent} at \eqref{saddle} is
\begin{equation}
    -\frac{32 i \sqrt{6}K^{3/2}L^3}{\pi^4}
\end{equation}
and we can apply the formula for quadratic fluctuation to deduce
\begin{align}
    |\mathcal{C}(2L, 0, 0, K)| &\approx 2 \left|\Re\left( \frac{1}{(2\pi i)^4}\frac{\sqrt{2\pi}^4}{\sqrt{32 i \sqrt{6}K^{3/2}L^3/\pi^4}}\exp(2\sqrt{2}\pi 24^{-1/4}K^{1/4}\sqrt{L})\right)\right|\nonumber\\
    &\approx 0.04K^{-3/4}L^{-3/2}\exp(4.0146K^{1/4}\sqrt{L})\label{asymptotic_with_quadratic_determinant}
\end{align}
The $1/(2\pi i)^4$ comes from the original $1/((2\pi)^4)$ outside the contour integral. The $\sqrt{2\pi}^4$ in the numerator comes from the four variables we need to integrate over. The denominator is simply the square root of the determinant of the Hessian at the saddle point.
Let us check this formula against numerical data. Take $2L = 50, K = 50${. The} logarithm of the exact result is
\begin{equation}
    \log(4773158006473089778) \approx 43
\end{equation}
The prediction is
\begin{equation}
    \log (0.04 50^{-3/4}25^{-3/2}) + 4.0146 50^{1/4}\sqrt{25}\approx 42.3959
\end{equation}
The error is within the predicted range: the next leading order term contributing the logarithm is of order $K^{2/4}$ when $K \approx 50$ it could contribute $\sqrt{50}\approx 7$ to the error. With some more work one can determine a more accurate constant prefactor (0.04 in \eqref{asymptotic_with_quadratic_determinant}) (\cite{NickRishiAndy} page 45)

\appendix

\section*{Appendices}
\section{Polylogarithms}
In this appendix we collect some basic facts about polylogarithms needed in this paper. The polylogarithm is defined as follows when $\Re(\theta)<0, s\in \Z_{>0}$:
\begin{equation}
    Li_s(\theta) = \sum_{n\geq 1}\frac{\exp(n\theta)}{n^s},
\end{equation}
This is an absolutely convergent series. Throughout this paper we require that $\Im(\theta)\neq 2\pi \Z$. This is the same as requiring $\exp(\theta)$ to be never real and positive. Our first goal is to perform analytic continuation from $\Re(\theta)<0$ to $\Re(\theta)\in \R$. First{, when} $s = 1$ we recognize it as the Taylor series of $-\log(1 - \exp(\theta))$:
\begin{equation}
    Li_1(\theta) = -\log(1 - \exp(\theta))
\end{equation}
(we choose the branch of log so that $\log(x)$ is real and positive when $x\in (1,\infty)$). Hence $Li_1(\theta)$ is analytic whenever $\exp(\theta)\not\in [1,\infty)$ which includes the whole domain $\Re(\theta)\in\R, \Im(\theta)\neq 2\pi \Z$. To perform analytic continuation for higher polylogarithms we use the identity $Li_s'(\theta) = Li_{s-1}(\theta), s\geq 2,\Re(\theta)<0$ and hence $Li_s(\theta)$ can be analytically continued to $\Re(\theta)\in\R$.

Next we collect the sum formula for $Li_s(\theta)$ valid for $0 < \Im(\theta) < 2\pi$. First we note
\begin{equation}
    Li_1(\theta) - Li_1(-\theta) = -\log\frac{1 - \exp(\theta)}{1 - \exp(-\theta)} = -\log(-\exp(\theta)) = -(\theta - i\pi)
\end{equation}
{Now we} keep using the identity $Li_s'(\theta) = Li_{s-1}(\theta), s\geq 2$ to deduce identities for higher polylogarithms. For example
\begin{equation}
    \frac{d}{d\theta}(Li_2(\theta) +Li_2(-\theta)) = \theta - i\pi\Rightarrow Li_2(\theta) + Li_2(-\theta) = -\frac{1}{2}(\theta - i\pi)^2 - \frac{\pi^2}{6}
\end{equation}
And similarly
\begin{eqnarray}
    Li_3(\theta) - Li_3( -\theta) = -\frac{1}{6}(\theta - i\pi)^3 - \frac{\pi^2}{6}\theta + \frac{i\pi^3}{6}\\
    Li_4(\theta) + Li_4( - \theta) = \frac{\pi^4}{45} - \frac{\theta^2(\theta-2\pi i)^2}{24}
\end{eqnarray}
We do not need higher polylogarithms in this paper.

\section{Estimates on the second and third derivative of F}
In this section we fill in a missing detail in "Step 5: Computation around the saddle points". We prove a bound on the third derivative of $F$ in \eqref{F_definition}: $\nabla^3F = \mathcal{O}_{K/L}(\log K)$ in the two balls $\tilde{B}^\pm$. The derivatives are taken with respect to $\sigma^i$. Using this bound one can prove that the Guassian integral approximation in the two balls are indeed valid and quadratic form in \eqref{gaussian_integral} tends to a constant. We will do the case of third partial derivative with respect to $\hat{m}$ and the other third partial derivatives can be analyzed using a similar method. We again split $F$ into four terms as in \eqref{four_terms}. Let us study the first term as an example, we want to prove the following:
\begin{equation}
    \partial^3_{\hat{m}} \sum_{n=1}^\infty\frac{1}{n} \frac{q_\tau^n\hat{y}^{-n}}{(1 - q_1^n)(1 - q_2^n)(1 - q_\tau^n)} = \mathcal{O}_{K/L}(K^{3/4}\log K)
\end{equation}
To do this we split the sum over $n$ into $n = 1... 1/(200\Re (\beta))$ and $n = 1/(200 \Re(\beta))...\infty$.
The estimate for the former is:
\begin{equation}
    \sum_{n=1}^{1/(200\Re(\beta))} = \frac{n^2q_\tau^n\hat{y}^{-n}}{(1 - q_1^n)(1 - q_2^n)(1 - q_\tau^n)} = \sum_{n=1}^{1/(200\Re(\beta))} \frac{n^2O(1)}{n\epsilon_1n\epsilon_2n\beta} = \mathcal{O}_{K/L}(K^{3/4}\log K)
\end{equation}
Here I have used the fact that $|q_\tau \hat{y}^{-1}|<1$ and $|(1 - \exp(-n\epsilon_1)| = \mathcal{O}_{K/L}( (n\epsilon_1)^{-1})$ whenever $|n\epsilon_1|<1$.
The estimate for the latter is:
\begin{equation}
    \sum_{n=1/(200\Re(\beta))}^{\infty} \frac{n^2q_\tau^n\hat{y}^{-n}}{(1 - q_1^n)(1 - q_2^n)(1 - q_\tau^n)} = \mathcal{O}_{K/L}(1) \sum_{n=1/(200\Re(\beta))}^{\infty}n^2 \exp(-(\Re(\beta) - |\Re \hat{m}|)^n = \mathcal{O}_{K/L}(K^{3/4})
\end{equation}
where I have use  the bound $|(1 -q_\tau^n)|^{-1} = \mathcal{O}_{K/L}(1)$ for this range of $n$ and one can explicitly evaluate the second infinite sum to obtain the desired inequality.

Next we show that the Hessian of $F$ at $\tilde{\sigma}^\pm$ converges to \eqref{hessian}. We will {compute} the $\hat{m}$ derivative first
\begin{equation}
    \partial^2_{\hat{m}}\sum_{n=1}^\infty \frac{1}{n}\frac{q_\tau^n\hat{y}^{-n}}{(1 - q_1^n)(1 - q_2^n)(1 - q_\tau^n)} = -\frac{\pi^2}{12\beta\epsilon_1\epsilon_2} + \mathcal{O}_{K/L}(K^{2/4}\log K)
\end{equation}
This implies that once we divide by $C(K,L)${,} the second derivative with respect to $\hat{m}$ at $\tilde{\sigma}^\pm$ converges to a finite limit as $K\to\infty$.
To prove this estimate we write
\begin{equation}
    \frac{1}{(1 - q_1^n)(1 - q_2^n)(1 - q_\tau^n)} = \frac{1}{n^3\beta\epsilon_1\epsilon_2} + g(n\epsilon_1,n\epsilon_2,n\beta)
\end{equation}
where the function $g$ satisfies $g(\epsilon_1,\epsilon_2,\beta) = \mathcal{O}_{K/L}(\epsilon_1^{-2})$ whenever $|\epsilon_1|<1,|\epsilon_2|<1, |\beta|<1$. We again split the sum into $n=1,...,1/(200 \Re(\beta))$ and $n = 1/(200 \Re(\beta)),...$ and in the former case we estimate
\begin{equation}
\sum_{n=1}^{1/(200\Re(\beta))} nq_\tau^n \hat{y}^{-n} g(n\epsilon_1,n\epsilon_2,n\beta) = \sum_{n=1}^{1/(200\Re(\beta))} \frac{q_\tau^n\hat{y}^{-n}}{n\epsilon_1^2} = \mathcal{O}_{K/L}(\log K\epsilon_1^{-2}) = \mathcal{O}_{K/L}(K^{2/4}\log K)
\end{equation}
In the latter case we estimate
\begin{equation}
    \sum_{1/(200\Re(\beta))}^\infty\frac{nq_\tau^n\hat{y}^{-n}}{(1 - q_1^n)(1 - q_2^n)(1 - q_\tau^n)} = \mathcal{O}_{K/L}(1) \sum_{1/(200\Re(\beta))}^\infty nq_\tau^n\hat{y}^{-n} = \mathcal{O}_{K/L}(K^{2/4})
\end{equation}
and
\begin{equation}
    \sum_{1/(200\Re(\beta))}^\infty \frac{q_\tau^n\hat{y}^{-n}}{n^2\beta\epsilon_1\epsilon_2} = \frac{\mathcal{O}_{K/L}(K^{-1/4})}{\beta\epsilon_1\epsilon_2} = \mathcal{O}_{K/L}(K^{2/4})
\end{equation}
Hence
\begin{equation}
    \partial^2_{\hat{m}}\sum_{n=1}^\infty \frac{1}{n}\frac{q_\tau^n\hat{y}^{-n}}{(1 - q_1^n)(1 - q_2^n)(1 - q_\tau^n)} = \sum_{n=1}^\infty \frac{q_\tau^n\hat{y}^{-n}}{n^2\beta\epsilon_1\epsilon_2} + \mathcal{O}_{K/L}(K^{2/4}\log K)
\end{equation}
Now we can rewrite the RHS using polylogarithms and deduce the desired estimate.
The analysis for the other three terms in \eqref{four_terms} and other derivatives are similar and we omit the details.

\section{Young tableaux}
In this appendix we set out our conventions for partitions and young tableaux. A partition $\lambda$ is a finite sequence of nonincreasing positive integers $\lambda_1\geq \lambda_2\geq \lambda_3... \geq \lambda_{l(\lambda)}>0$ where $l(\lambda)$ is called the length of the partition. One usually visualizes a partition by drawing the associated Young diagram $Y(\lambda)$. The young diagram is made up of adjacent boxes. The top row of the young diagram has $\lambda_1$ boxes, the second row has $\lambda_2$ boxes, and the $p$th row has $\lambda_p$ boxes. Henceforth any operation on a young diagram induces an operation on the associated partition and vice versa. A box $s$ in a young diagram $Y(\lambda)$ is labelled by its cooordinates $(p,q)$ where $p = 1,..., l(\lambda)$ is the vertical coordinate and $q = 1,..., \lambda_p$ is the horizontal coordinate. The arm length $a(p,q)$ of a box $s = (p,q)$ is  defined as the number of boxes to the right of $s$. The leg length $l(p,q)$ of $s$ is the number of boxes below $s$.
\begin{equation}\label{leg_arm_length}
    a(s) = a(p,q) = \lambda_p - q, l(s) = l(p,q) = \lambda^T_q - p
\end{equation}
where the transpose of a partition is the reflection of the partition by the 45 degree axis from top left to bottom right. For example the transpose of $\lambda_1 = 3,\lambda_2 = 2,\lambda_3 = 2$ is the partition $\lambda_1 = 3, \lambda_2 = 3,\lambda_3 = 1$.
The length of the transposed partition is given by $l(\lambda^T) = \lambda_1$. 

\section*{Acknowledgements}
We thank Rishi Mouland, Canberk Sanli, Robert Osburn, Mark Gross, Samuel Crew for useful discussions. BZ would like to thank Irina Davydenkova for her generous assistance with parrallel computing. BZ is supported by a Trinity College internal studentship. The research of BZ and ND are supported by STFC consolidated grants ST/T000694/1 and ST/X000664/1.

\bibliographystyle{unsrt}
\bibliography{Refs}

\begin{thebibliography}{10}

\bibitem{Witten}
Edward Witten.
\newblock {Supersymmetry and Morse theory}.
\newblock {\em J. Diff. Geom.}, 17(4):661--692, 1982.

\bibitem{Eliezer}
S.~Fubini and E.~Rabinovici.
\newblock {SUPERCONFORMAL QUANTUM MECHANICS}.
\newblock {\em Nucl. Phys. B}, 245:17, 1984.

\bibitem{Ivanov}
Sergey Fedoruk, Evgeny Ivanov, and Olaf Lechtenfeld.
\newblock {Superconformal Mechanics}.
\newblock {\em J. Phys. A}, 45:173001, 2012.

\bibitem{Singleton1}
Andrew Singleton.
\newblock {Superconformal quantum mechanics and the exterior algebra}.
\newblock {\em JHEP}, 06:131, 2014.

\bibitem{SingletonDorey}
Nick Dorey and Andrew Singleton.
\newblock An index for superconformal quantum mechanics.
\newblock {\em arXiv preprint arXiv:1812.11816}, 2018.

\bibitem{braden2012}
Tom Braden, Nicholas Proudfoot, and Ben Webster.
\newblock Quantizations of conical symplectic resolutions i: local and global structure.
\newblock {\em arXiv preprint arXiv:1208.3863}, 2012.

\bibitem{DoreyBarnsGraham}
Alec~E. Barns-Graham and Nick Dorey.
\newblock {A Superconformal Index for HyperK\"ahler Cones}.
\newblock {\em arXiv preprint arXiv:1812.04565}, 2018.

\bibitem{NickRishiAndy}
Nick Dorey, Rishi Mouland, and Boan Zhao.
\newblock {Black Hole Entropy from Quantum Mechanics}.
\newblock {\em arXiv preprint arXiv 2207.12477}, 2022.

\bibitem{LeeNahmgoong}
Kimyeong Lee and June Nahmgoong.
\newblock {Cardy Limits of 6d Superconformal Theories}.
\newblock {\em JHEP}, 05:118, 2021.

\bibitem{kim2017asymptoticM5brane}
Seok Kim and June Nahmgoong.
\newblock {Asymptotic M5-brane entropy from S-duality}.
\newblock {\em JHEP}, 12:120, 2017.

\bibitem{AlvarezGaumePotential}
Luis Alvarez-Gaume and Daniel~Z. Freedman.
\newblock {Potentials for the Supersymmetric Nonlinear Sigma Model}.
\newblock {\em Commun. Math. Phys.}, 91:87, 1983.

\bibitem{Rains}
Eric~M. Rains and S.~Ole Warnaar.
\newblock A nekrasov{\textendash}okounkov formula for macdonald polynomials.
\newblock {\em Journal of Algebraic Combinatorics}, 48(1):1--30, sep 2017.

\bibitem{HardyRamanujan}
G.~H. Hardy and S.~Ramanujan.
\newblock {Asymptotic Formulae in Combinatory Analysis}.
\newblock {\em Proceedings of the London Mathematical Society}, s2-17(1):75--115, 01 1918.

\bibitem{SingletonThesis}
Andrew Singleton.
\newblock {The Geometry and Representation Theory of Superconformal Quantum Mechanics}.
\newblock \url{https://www.repository.cam.ac.uk/handle/1810/260821}.

\bibitem{nakajima1999lectures}
H.~Nakajima and American~Mathematical Society.
\newblock {\em Lectures on Hilbert Schemes of Points on Surfaces}.
\newblock University lecture series. American Mathematical Society, 1999.

\bibitem{Braverman}
Maxim Braverman.
\newblock Background cohomology of a non-compact kahler g-manifold, 2012.

\bibitem{AlvarezGaumeIndex}
Luis Alvarez-Gaume.
\newblock {Supersymmetry and the Atiyah-Singer Index Theorem}.
\newblock {\em Commun. Math. Phys.}, 90:161, 1983.

\bibitem{Nakajima}
Hiraku Nakajima and Kota Yoshioka.
\newblock {Instanton counting on blowup. 1.}
\newblock {\em Invent. Math.}, 162:313--355, 2005.

\bibitem{Newman}
D.~J. Newman.
\newblock {A simplified proof of the partition formula.}
\newblock {\em Michigan Mathematical Journal}, 9(3):283 -- 287, 1962.

\end{thebibliography}

\end{document}